\documentclass[aps,nofootinbib,preprint,superscriptaddress]{revtex4}%
\usepackage{hyperref}
\usepackage{amsmath}
\usepackage{amsfonts}
\usepackage{amssymb}
\usepackage{graphicx}
\usepackage{color}%
\providecommand{\U}[1]{\protect\rule{.1in}{.1in}}

\begin{document}
\title{Stringy scaling of multi-tensor hard string scattering amplitudes and the $K$-identities}
\author{Sheng-Hong Lai}
\email{xgcj944137@gmail.com}
\affiliation{Department of Electrophysics, National Yang Ming Chiao Tung University,
Hsinchu, Taiwan, R.O.C.}
\author{Jen-Chi Lee}
\email{jcclee@cc.nctu.edu.tw}
\affiliation{Department of Electrophysics, National Yang Ming Chiao Tung University,
Hsinchu, Taiwan, R.O.C.}
\affiliation{Center for Theoretical and Computational Physics (CTCP), National Yang Ming
Chiao Tung University, Hsinchu, Taiwan, R.O.C.}
\author{Yi Yang}
\email{yiyang@mail.nctu.edu.tw}
\affiliation{Department of Electrophysics, National Yang Ming Chiao Tung University,
Hsinchu, Taiwan, R.O.C.}
\affiliation{Center for Theoretical and Computational Physics (CTCP), National Yang Ming
Chiao Tung University, Hsinchu, Taiwan, R.O.C.}
\affiliation{National Center for Theoretical Physics, R.O.C.}
\date{\today}

\begin{abstract}
We calculate $n$-point hard string scattering amplitudes ($HSSA$) with $n-2$
tachyons and $2$ tensor states at arbitrary mass levels. We discover the
stringy scaling behavior of these $HSSA$. It is found that for $HSSA$ with
more than $2$ transverse directions, the degree of stringy scaling
dim$\mathcal{M}_{2}$ decreases comparing to the degree of stringy scaling
dim$\mathcal{M}_{1}$ of the $n-1$ tachyons and $1$ tensor $HSSA$ calculated
previously. Moreover, we propose a set of $K$-identities\ which is the key to
demonstrate the stringy scaling behavior of $HSSA$. We explicitly prove both
the diagonal and off-diagonal $K$-identities for the $4$-point $HSSA$ and give
numerical proofs of these $K$-identities for some higher point $HSSA$.

\end{abstract}
\maketitle

\bigskip\setcounter{equation}{0} \renewcommand{\theequation}{\arabic{section}.\arabic{equation}}

\section{Introduction}

Recently, it was discovered \cite{Regge, hard,Komaba} that the degree of
freedom or number of kinematic variables reduced for the $n$-point with $n-1$
tachyons and $1$ tensor hard string scattering amplitudes ($HSSA$) and the
Regge string scattering amplitudes ($RSSA$). Historically, the first
\textit{stringy scaling} behavior was conjectured by Gross
\cite{GM,GM1,Gross,GrossManes} for the $4$-point $HSSA$ and later explicitly
proved by Taiwan group in \cite{ChanLee,ChanLee2,CHLTY2,CHLTY1}. Indeed, all
functional forms of $4$-point $HSSA$ at each fixed mass level are found to be
proportional to each other with \textit{constant} ratios (independent of the
scattering angle $\phi$, or the deficit of the kinematics variable
dim$\mathcal{M}_{1}=1-0=1$). See the recent reviews \cite{review, over}.

This stringy scaling behavior, similar to the $QCD$ Bjorken scaling
\cite{bs,cg} and its $GLAP$ correction \cite{GL,AP}, was recently generalized
to arbitrary $n$-point ($n-1$ tachyons and $1$ tensor) $HSSA$ with $n\geq5$
\cite{Regge,hard,loop}. As an example \cite{hard}, the number of degree of
freedom of $6$-point $HSSA$ ($5$ tachyons and $1$ tensor) with $3$ transverse
directions reduces from $8$ to $2$, and dim$\mathcal{M}_{1}=8-2=6$.

Since massive higher spin tensor string states receive more degree of freedom
of spin polarizations, it is tempted to speculate that the degree of stringy
scaling or dim$\mathcal{M}_{k}$ ($k\leq n$) for the $n$-point $k$-tensor
$HSSA$ with $k\geq2$ may decrease and is smaller than dim$\mathcal{M}_{1}$. In
this paper, we calculate $n$-point ($n\geq4$) $HSSA$ with $n-2$ tachyons and
$2$ high energy tensor states at arbitrary mass levels. It is found that the
degree of stringy scaling dim$\mathcal{M}_{2}$ for $HSSA$ with more than $2$
transverse directions or $r\geq3$ (see Eq.(\ref{rr})) indeed decreases
comparing to that of the $1$ tensor $HSSA$ dim$\mathcal{M}_{1}$ calculated previously.

In the saddle-point calculation of the $HSSA$, the saddle-point of the
integration can be exactly solved only for some lower point $HSSA$
calculation. For the cases of general $n$-point saddle-point $HSSA$
calculation, we propose a set of the so-called $K$-identities. We will see
that without directly solving the saddle-point equation, these $K$-identities
can be used to simplify the $HSSA$ calculation, and turn out to be the key
step of the calculation to demonstrate the stringy scaling behavior.

Some of these $K$-identities were originally proposed for the calculation of
$n$-point $1$ tensor $HSSA$ \cite{Regge,hard}. They are the direct results of
the identification of two different calculation of $HSSA$, namely, the
saddle-point calculation and the decoupling of zero-norm state ($ZNS$)
calculation \cite{review, over}. However, only $K$-identities for the
$4$-point $HSSA$ with $1$ and $2$ tensors will be analytically proved in this
paper, other $K$-identities of higher point ($n\geq5$) $HSSA$ calculation will
be tested numerically. Presumably, these $K$-identities are results of
conservation of momenta in the hard scattering limit \textit{at the
saddle-point }\cite{KID}.

For the saddle-point calculation of $n$-point $2$ tensor $HSSA$, in addition
to the diagonal $K$-identities one obtained during the calculation of
$n$-point $1$ tensor $HSSA$, it turns out that one needs to introduce $1$ more
off-diagonal $K$-identity. This off-diagonal $K$-identity will contribute to
the calculation of degree of stringy scaling dim$\mathcal{M}_{2}$ for $HSSA$
out of the scattering plane or $r\geq2$. It is conjectured that for general
higher $k$ tensor $HSSA$ calculation ($k\leq n$), these off-diaganal
$K$-identities are crucial to show the consistent result that dim$\mathcal{M}%
_{k}\geq0$.

This paper is organized as following. In section II, we first calculate an
example of $4$-point $2$ tensor $HSSA$ for mass level $M^{2}=2$. We then
extend the calculation to arbitrary mass levels and obtain the ratios
consistent with calculation from the method of decoupling of $ZNS$. In section
III, we calculate general $n$-point $2$ tensor $HSSA$. One key step is to
propose both the diagonal and off-diagonal $K$-identities to simplify the
calculation and obtain the consistent ratios calculated from the method of
$ZNS$. In section IV, we discuss in details the $K$-identities and apply them
to the calculation of the degree of stringy scaling dim$\mathcal{M}_{2}$. We
also explicitly verify all $4$-point $1$ and $2$ tensors $K$-identities and
discuss the numerical check of the higher point $K$-identities. A brief
conclusion is given in section V.

\setcounter{equation}{0}

\section{The four-point two tensor $HSSA$}

In this section, we will use the saddle-point method to calculate the
$4$-point $HSSA$ with $2$ tachyons and $2$ tensors. We begin with a brief
review of the calculation of $HSSA$ with $3$ tachyons and $1$ tensor state
\cite{ChanLee,ChanLee2,CHLTY2,CHLTY1}. For the $4$-point $HSSA$, we first note
that in the hard scattering limit, components of polarization orthogonal to
the scattering plane are subleading order in energy. Defining $e^{P}=\frac
{1}{M_{2}}(E_{2},\mathrm{k}_{2},\vec{0})=\frac{k_{2}}{M_{2}}$ the momentum
polarization, $e^{L}=\frac{1}{M_{2}}(\mathrm{k}_{2},E_{2},\vec{0})$ the
longitudinal polarization and the transverse polarization $e^{T}=(0,0,1)$ on
the scattering plane, it can then be shown that at each fixed mass level
$M^{2}=2(N-1)$ only states of the following form
\cite{ChanLee,ChanLee2,CHLTY2,CHLTY1} (in the hard scattering limit
$e^{P}\simeq$ $e^{L}$)
\begin{equation}
\left\vert N,2m,q\right\rangle =\left(  \alpha_{-1}^{T}\right)  ^{N-2m-2q}%
\left(  \alpha_{-1}^{L}\right)  ^{2m}\left(  \alpha_{-2}^{L}\right)
^{q}\left\vert 0;k\right\rangle \label{11}%
\end{equation}
are leading order in energy. Note in particular that states with $\left(
\alpha_{-1}^{L}\right)  ^{2m+1}$ are of subleading order in energy. To achieve
physical state condition in the old covariant first quantized string, it can
be shown that high energy string states in the physical state basis ($PSB$)
with Virasoro constraints imposed can be written as a linear combination of
states in the oscillator state basis ($OSB$) in Eq.(\ref{11}) \cite{review,
over}.

There are two different methods to calculate the stringy scaling behavior of
$HSSA$, the method of decoupling of $ZNS$ and the saddle-point method. Both
methods lead to\textbf{ }constant ratios among $4$-point $HSSA$ at each fixed
mass level \cite{ChanLee,ChanLee2,CHLTY2,CHLTY1}%
\begin{equation}
\frac{A^{\left(  N,2m,q\right)  }}{A^{\left(  N,0,0\right)  }}=\frac{\left(
2m\right)  !}{m!}\left(  \frac{-1}{2M}\right)  ^{2m+q}%
.\text{(\textbf{independent of }}\phi\text{ !).} \label{22}%
\end{equation}
In Eq.(\ref{22}), $A^{\left(  N,m,q\right)  }$ is the $4$-point $HSSA$ of any
string vertex $V_{j}$ with $j=1,3,4$ and $V_{2}$ is the high energy state in
Eq.(\ref{11}). $A^{\left(  N,0,0\right)  }$ is the $4$-point $HSSA$ of any
string vertex $V_{j}$ with $j=1,3,4$, and $V_{2}$ is the leading Regge
trajectory string state at mass level $N$. Note that we have omitted the
tensor indice of $V_{j}$ with $j=1,3,4$ and keep only those of $V_{2}$ in
$\mathcal{T}^{\left(  N,2m,q\right)  }$.

For the $j$ tensor cases with $j=2,3$ and $4,$ it is easy to argue from the
method of the decoupling of $ZNS$ that the ratios are $j$ products of ratio in
Eq.(\ref{22}). However, it is highly nontrivial to calculate $n$-point
multi-tensor $HSSA$ at arbitrary mass levels. Moreover, as we will see in
section IV that one important step of this saddle-point calculation is to
introduce the $K_{i}$ vectors and $K$-identities, and give information to
calculate the degree of stringy scaling dim$\mathcal{M}_{k}$.

For later reference, we begin with a brief description of the $4$-point
saddle-point calculation with $3$ tachyons and $1$ tensor state here. The
$t-u$ channel of $HSSA$ with $3$ tachyons and $1$ high energy state in
Eq.(\ref{11}) can be calculated to be (after $SL(2,R)$ fixing)%
\begin{align}
A^{(N,2m,q)}  &  =\int_{1}^{\infty}dzz^{(1,2)}(1-z)^{(2,3)}\left[  \frac
{e^{T}\cdot k_{1}}{z}-\frac{e^{T}\cdot k_{3}}{1-z}\right]  ^{N-2m-2q}%
\nonumber\\
&  \cdot\left[  \frac{e^{P}\cdot k_{1}}{z}-\frac{e^{P}\cdot k_{3}}%
{1-z}\right]  ^{2m}\left[  -\frac{e^{P}\cdot k_{1}}{z^{2}}-\frac{e^{P}\cdot
k_{3}}{(1-z)^{2}}\right]  ^{q}%
\end{align}
where $(1,2)=k_{1}\cdot k_{2}$ etc. We can rewrite the amplitude above into
the following form%
\begin{equation}
A^{(N,2m,q)}(K)=\int_{1}^{\infty}dz\mbox{ }u(z)e^{-\Lambda f(z)} \label{int}%
\end{equation}
where
\begin{align}
\Lambda &  \equiv-(1,2)\rightarrow\frac{s}{2}\rightarrow2E^{2},\\
\tau &  \equiv-\frac{(2,3)}{(1,2)}\rightarrow-\frac{t}{s}\rightarrow\sin
^{2}\frac{\phi}{2},\\
f(z)  &  \equiv\ln z-\tau\ln(1-z),\label{f}\\
u(z)  &  \equiv\left[  \frac{(1,2)}{M}\right]  ^{2m+q}(1-z)^{-N+2m+2q}%
\underset{\ast}{\underbrace{(f^{\prime})^{2m}}}(f^{\prime\prime})^{q}%
(-e^{T}\cdot k_{3})^{N-2m-2q}.
\end{align}
The saddle-point $z=\tilde{z}$, is defined by
\begin{equation}
f^{\prime}(\tilde{z})=0,
\end{equation}
which can be exactly solved to be%
\begin{equation}
\tilde{z}=\frac{1}{1-\tau}=\sec^{2}\frac{\phi}{2}. \label{saddle}%
\end{equation}
It is now easy to see that%
\begin{equation}
u(\tilde{z})=u^{\prime}(\tilde{z})=....=u^{(2m-1)}(\tilde{z})=0, \label{even}%
\end{equation}
and
\begin{equation}
u^{(2m)}(\tilde{z})=\left[  \frac{(1,2)}{M}\right]  ^{2m+q}(1-\tilde
{z})^{-N+2m+2q}(2m)!(f_{0}^{\prime\prime})^{2m+q}(-e^{T}\cdot k_{3})^{N-2m-2q}
\label{aa}%
\end{equation}
where $f_{0}^{\prime\prime}=f^{\prime\prime}(\tilde{z})$.

It is important to note that in calculating $HSSA$ containing high energy
states in Eq.(\ref{11}) with $\left(  \alpha_{-1}^{L}\right)  ^{2m}$, $m\neq
0$, the \textit{naive} energy order of the amplitudes based on the naive mass
counting vanishes \cite{ChanLee,ChanLee2,CHLTY2,CHLTY1}, and the real leading
energy order $HSSA$ will in general drop by an even number of energy power as
can be seen from Eq.(\ref{even}). On the other hand, $HSSA$ corresponding to
states with $(\alpha_{-1}^{L})^{2m+1}$\ turn out to be of subleading order in energy.

With these inputs, one can evaluate the Gaussian integral in Eq.(\ref{int})
and show explicitly that the angular dependence of $HSSA$ depends only on the
level $N$ but not $m$ and $q$%
\begin{equation}
A^{(N,2m,q)}=\sqrt{\frac{2\pi}{Kf_{0}^{\prime\prime}}}e^{-\Lambda f_{0}%
}\left[  (-1)^{N-q}\frac{2^{N-2m-q}(2m)!}{m!\ {M}^{2m+q}}\ \tau^{-\frac{N}{2}%
}(1-\tau)^{\frac{3N}{2}}E^{N}+O(E^{N-2})\right]  ,
\end{equation}
and we have \cite{ChanLee,ChanLee2,CHLTY2,CHLTY1}
\begin{align}
\lim_{E\rightarrow\infty}\frac{A^{(N,2m,q)}}{A^{(N,0,0)}}  &  =\frac
{(-1)^{q}(2m)!}{m!(2M)^{2m+q}}=(-\frac{2m-1}{M})....(-\frac{3}{M})(-\frac
{1}{M})(-\frac{1}{2M})^{m+q}\nonumber\\
&  =\left(  \frac{1}{2}\right)  ^{m+q}\left(  -\frac{1}{M}\right)
^{q+2m}\left(  2m-1\right)  !!,
\end{align}
which is\textbf{ }remarkably\textbf{ }consistent with calculation of
decoupling of high energy $ZNS$ obtained in Eq.(\ref{22}).

We now turn to calculate the $4$-point $HSSA$ with $2$ tachyons and $2$ tensor
states. We choose to put the $2$ tensors on $z_{2}$ and $z_{3}$. The
kinematics of $4$-points $HSSA$ can be written as%

\begin{align}
k_{1}  &  =%
\begin{pmatrix}
E\text{,} & -E\text{,} & 0
\end{pmatrix}
,\nonumber\\
k_{2}  &  =%
\begin{pmatrix}
E\text{,} & E\text{,} & 0
\end{pmatrix}
,\nonumber\\
k_{3}  &  =%
\begin{pmatrix}
-E\text{,} & -E\cos\phi\text{,} & -E\sin\phi
\end{pmatrix}
,\nonumber\\
k_{4}  &  =%
\begin{pmatrix}
-E\text{,} & E\cos\phi\text{,} & E\sin\phi
\end{pmatrix}
,
\end{align}
which imply%
\begin{align}
k_{12}  &  =-2E^{2},\nonumber\\
k_{13}  &  =E^{2}\left(  1-\cos\phi\right)  ,\nonumber\\
k_{23}  &  =E^{2}\left(  1+\cos\phi\right)  .
\end{align}
For the $2$ tensor cases, we will use the following short notations for
polarizations
\begin{align}
T_{2}  &  =%
\begin{pmatrix}
0\text{,} & 0\text{,} & 1
\end{pmatrix}
,\nonumber\\
T_{3}  &  =%
\begin{pmatrix}
0\text{,} & -\sin\phi\text{,} & \cos\phi
\end{pmatrix}
,\nonumber\\
L_{2}  &  =\frac{1}{M_{2}}%
\begin{pmatrix}
E\text{,} & E\text{,} & 0
\end{pmatrix}
,\nonumber\\
L_{3}  &  =\frac{1}{M_{3}}%
\begin{pmatrix}
-E\text{,} & -E\cos\phi\text{,} & -E\sin\phi
\end{pmatrix}
\end{align}
where $T_{2}$ means $e^{T_{2}}$ etc. One can easily calculate the following
projections
\begin{equation}
k_{1}^{T_{2}}=0,\text{ }k_{3}^{T_{2}}=-E\sin\phi,\text{ }k_{1}^{T_{3}}%
=E\sin\phi,\text{ }k_{2}^{T_{3}}=-E\sin\phi;
\end{equation}%
\begin{equation}
k_{1}^{L_{2}}=\frac{-2E^{2}}{M_{2}},\text{ }k_{3}^{L_{2}}=\frac{2E^{2}}{M_{2}%
}\tau,\text{ }k_{1}^{L_{3}}=\frac{2E^{2}}{M_{3}}\left(  1-\tau\right)  ,\text{
}k_{2}^{L_{3}}=\frac{2E^{2}}{M_{3}}\tau;
\end{equation}
and after $SL(2,R)$ fixing%
\begin{equation}
f_{0}^{\prime\prime}=-\frac{1}{\tilde{z}^{2}}+\frac{\tau}{\left(  1-\tilde
{z}\right)  ^{2}}=\frac{\left(  1-\tau\right)  ^{3}}{\tau}.
\end{equation}
\setcounter{equation}{0}

\subsection{An example of mass level $M^{2}=2$}

As a simple example of $4$-point $HSSA$ with $2$ tachyons and $2$ tensor
states, we first calculate $HSSA$ for tensor states at mass level $N=2$. By
Eq.(\ref{11}), there are $3$ high energy states in the $OSB$%
\begin{equation}
\left(  \alpha_{-1}^{T}\right)  ^{2}\left\vert 0;k\right\rangle ,\text{
}\left(  \alpha_{-2}^{L}\right)  \left\vert 0;k\right\rangle ,\text{ }\left(
\alpha_{-1}^{L}\right)  ^{2}\left\vert 0;k\right\rangle . \label{t}%
\end{equation}
By Eq.(\ref{22}), the ratios of $HSSA$ with $3$ tachyons and $1$ tensor in
Eq.(\ref{t}) are
\begin{equation}
A^{T_{2}T_{2}}:A^{L_{2}}:A^{L_{2}L_{2}}=4:\sqrt{2}:1. \label{4}%
\end{equation}
For the case of \ $HSSA$ with $2$ tachyons and $2$ tensors, the calculation of
the decoupling of zero-norm state ($ZNS)$ gives
\begin{align}
&  \left(  T_{2}T_{2}\oplus L_{2}\oplus L_{2}L_{2}\right)  \otimes\left(
T_{3}T_{3}\oplus L_{3}\oplus L_{3}L_{3}\right) \nonumber\\
&  =\left(  4:\sqrt{2}:1\right)  \times\left(  4:\sqrt{2}:1\right)  .
\label{44}%
\end{align}

We will recalculate the ratios by using the saddle-point method. In
calculating the $HSSA$, one needs to pay attention to the state $\left(
\alpha_{-1}^{L}\right)  ^{2}\left\vert 0;k\right\rangle $ with $m=1$ in
Eq.(\ref{t}), as has been mentioned in the paragraph after Eq.(\ref{aa}). The
calculation for $HSSA$ with $m_{2}=m_{3}=0$ are straightforward and we obtain
\begin{align}
A^{T_{2}T_{2};T_{3}T_{3}}  &  =\int d^{4}ze^{-\Lambda f}\left(  \frac
{k_{1}^{T_{2}}}{z_{1}-z_{2}}+\frac{k_{3}^{T_{2}}}{z_{3}-z_{2}}+\frac
{k_{4}^{T_{2}}}{z_{4}-z_{2}}\right)  ^{2}\left(  \frac{k_{1}^{T_{3}}}%
{z_{1}-z_{3}}+\frac{k_{2}^{T_{3}}}{z_{2}-z_{3}}+\frac{k_{4}^{T_{3}}}%
{z_{4}-z_{3}}\right)  ^{2}\nonumber\\
&  =e^{-\Lambda f_{0}}\left(  4E^{2}\frac{\left(  1-\tau\right)  ^{3}}{\tau
}\right)  \left(  4E^{2}\frac{1-\tau}{\tau}\right)  \sqrt{\frac{2\pi}%
{f_{0}^{\prime\prime}}}, \label{111}%
\end{align}%
\begin{align}
A^{T_{2}T_{2};L_{3}}  &  =\int d^{4}ze^{-\Lambda f}\left(  \frac{k_{1}^{T_{2}%
}}{z_{1}-z_{2}}+\frac{k_{3}^{T_{2}}}{z_{3}-z_{2}}+\frac{k_{4}^{T_{2}}}%
{z_{4}-z_{2}}\right)  ^{2}\left(  \frac{k_{1}^{L_{3}}}{\left(  z_{1}%
-z_{3}\right)  ^{2}}+\frac{k_{2}^{L_{3}}}{\left(  z_{2}-z_{3}\right)  ^{2}%
}+\frac{k_{4}^{L_{3}}}{\left(  z_{4}-z_{3}\right)  ^{2}}\right) \nonumber\\
&  =e^{-\Lambda f_{0}}\left(  4E^{2}\frac{\left(  1-\tau\right)  ^{3}}{\tau
}\right)  \left(  \frac{2E^{2}\left(  1-\tau\right)  }{M_{3}\tau}\right)
\sqrt{\frac{2\pi}{f_{0}^{\prime\prime}}}. \label{222}%
\end{align}
With $M_{3}=\sqrt{2}$, Eq.(\ref{111}) and Eq.(\ref{222}) give the consistent
ratios in Eq.(\ref{4}).

We now turn to calculate $HSSA$ containing state $\left(  \alpha_{-1}%
^{L}\right)  ^{2}\left\vert 0;k\right\rangle $ with $m_{2}=0$, $m_{3}=1$. The
saddle-point calculation gives
\begin{align}
A^{T_{2}T_{2};L_{3}L_{3}}  &  =\int d^{4}ze^{-\Lambda f}\left(  \frac
{k_{1}^{T_{2}}}{z_{1}-z_{2}}+\frac{k_{3}^{T_{2}}}{z_{3}-z_{2}}+\frac
{k_{4}^{T_{2}}}{z_{4}-z_{2}}\right)  ^{2}\left(  \frac{k_{1}^{L_{3}}}%
{z_{1}-z_{3}}+\frac{k_{2}^{L_{3}}}{z_{2}-z_{3}}+\frac{k_{4}^{L_{3}}}%
{z_{4}-z_{3}}\right)  ^{2}\nonumber\\
&  =\int dz_{2}e^{-\Lambda f}\left(  -E\sin\phi\frac{\tau-1}{\tau}\right)
^{2}\left(  \frac{\frac{2E^{2}}{M_{3}}\cos^{2}\frac{\phi}{2}}{-1}+\frac
{\frac{2E^{2}}{M_{3}}\sin^{2}\frac{\phi}{2}}{z_{2}-1}\right)  ^{2}.
\end{align}
We note that the above integrand vanish if we evaluate $z_{2}$ at the saddle
point $z_{2}$ $=\tilde{z}_{2}$. To obtain the real energy order $HSSA$, one
encounters the naive energy order problem and we need to do Tayor expansion
about the saddle point $\tilde{z}_{2}$ before performing the integration. By
doing so, we obtain
\begin{align}
A^{T_{2}T_{2};L_{3}L_{3}}  &  =\int dz_{2}e^{-\Lambda f}\left(  -2E\sqrt
{\tau\left(  1-\tau\right)  }\frac{\tau-1}{\tau}+\cdots\right)  ^{2}%
\nonumber\\
&  \cdot\left(  -\frac{2E^{2}}{M_{3}}\left(  1-\tau\right)  +\frac{2E^{2}%
}{M_{3}}\tau\frac{1-\tau}{\tau}-\frac{\frac{2E^{2}}{M_{3}}\sin^{2}\frac{\phi
}{2}}{\left(  z_{20}-1\right)  ^{2}}\left(  z_{2}-\tilde{z}_{2}\right)
+\cdots\right)  ^{2}\nonumber\\
&  =\int dz_{2}e^{-\Lambda f}\left(  -2E\sqrt{\tau\left(  1-\tau\right)
}\frac{\tau-1}{\tau}+\cdots\right)  ^{2}\left(  0-\frac{\frac{2E^{2}}{M_{3}%
}\sin^{2}\frac{\phi}{2}}{\left(  z_{20}-1\right)  ^{2}}\left(  z_{2}-\tilde
{z}_{2}\right)  +\cdots\right)  ^{2}\nonumber\\
&  =e^{-\Lambda f_{0}}\left(  4E^{2}\frac{\left(  1-\tau\right)  ^{3}}{\tau
}\right)  \left(  \frac{2E^{2}}{M_{3}^{2}}\frac{\left(  1-\tau\right)  }{\tau
}\right)  \sqrt{\frac{2\pi}{f_{0}^{\prime\prime}}},
\end{align}
which gives the ratio consistent with Eq.(\ref{4}). Similar calculation can be
performed for other $HSSA$ in Eq.(\ref{4}) and all consistent ratios can be
obtained except the $A^{L_{2}L_{2};L_{3}L_{3}}$ one with $m_{2}=m_{3}=1$ which
needs to be treated carefully.

We first note that in calculating $HSSA$ in Eq.(\ref{44}), contraction terms
from $\partial_{2}X_{2}\cdot\cdot$ and $\partial_{3}X_{3}\cdot\cdot$ are
considered to be of subleading order in energy and can be ignored. However,
for the special case of $A^{L_{2}L_{2};L_{3}L_{3}}$ amplitude, one encounters
the \textit{double naive energy order problem}, and it turns out that one has
to include contraction terms from $\partial_{2}X_{2}\cdot\cdot$ and
$\partial_{3}X_{3}\cdot\cdot$ in order to achieve the consistent ratios
calculated by the method of decoupling of $ZNS$ in Eq.(\ref{44}).

We thus follow the standard $KLT$ \cite{KLT} method to calculate the
amplitude
\begin{align}
A^{L_{2}L_{2};L_{3}L_{3}} &  =\int d^{4}ze^{ik_{1}X_{1}}e^{ik_{2}X_{2}%
+i\xi_{2}\partial_{2}X_{2}+i\xi_{2}^{\prime}\partial_{2}X_{2}}e^{ik_{3}%
X_{3}+i\xi_{3}\partial_{3}X_{3}+i\xi_{3}^{\prime}\partial_{3}X_{3}}%
e^{ik_{4}X_{4}}\\
&  =\int d^{4}ze^{-\Lambda f}\exp\left\{
\begin{array}
[c]{c}%
\left[  \frac{k_{1}\cdot\xi_{2}}{z_{1}-z_{2}}+\frac{k_{3}\cdot\xi_{2}}%
{z_{3}-z_{2}}\right]  +\left[  \frac{k_{1}\cdot\xi_{2}^{\prime}}{z_{1}-z_{2}%
}+\frac{k_{3}\cdot\xi_{2}^{\prime}}{z_{3}-z_{2}}\right]  \\
+\left[  \frac{k_{1}\cdot\xi_{3}}{z_{1}-z_{3}}+\frac{k_{2}\cdot\xi_{3}}%
{z_{2}-z_{3}}\right]  +\left[  \frac{k_{1}\cdot\xi_{3}^{\prime}}{z_{1}-z_{3}%
}+\frac{k_{2}\cdot\xi_{3}^{\prime}}{z_{2}-z_{3}}\right]  \\
+\left[  \frac{\xi_{2}\cdot\xi_{3}}{\left(  z_{3}-z_{2}\right)  ^{2}}%
+\frac{\xi_{2}\cdot\xi_{3}^{\prime}}{\left(  z_{3}-z_{2}\right)  ^{2}}%
+\frac{\xi_{2}^{\prime}\cdot\xi_{3}}{\left(  z_{3}-z_{2}\right)  ^{2}}%
+\frac{\xi_{2}^{\prime}\cdot\xi_{3}^{\prime}}{\left(  z_{3}-z_{2}\right)
^{2}}\right]
\end{array}
\right\}  _{multilinear}\\
&  =\int d^{4}ze^{-\Lambda f}\left\{
\begin{array}
[c]{c}%
\left[  \frac{k_{1}\cdot\xi_{2}}{z_{1}-z_{2}}+\frac{k_{3}\cdot\xi_{2}}%
{z_{3}-z_{2}}\right]  \left[  \frac{k_{1}\cdot\xi_{2}^{\prime}}{z_{1}-z_{2}%
}+\frac{k_{3}\cdot\xi_{2}^{\prime}}{z_{3}-z_{2}}\right]  \left[  \frac
{k_{1}\cdot\xi_{3}}{z_{1}-z_{3}}+\frac{k_{2}\cdot\xi_{3}}{z_{2}-z_{3}}\right]
\left[  \frac{k_{1}\cdot\xi_{3}^{\prime}}{z_{1}-z_{3}}+\frac{k_{2}\cdot\xi
_{3}^{\prime}}{z_{2}-z_{3}}\right]  \\
+\frac{\xi_{2}\cdot\xi_{3}}{\left(  z_{3}-z_{2}\right)  ^{2}}\left[
\frac{k_{1}\cdot\xi_{2}^{\prime}}{z_{1}-z_{2}}+\frac{k_{3}\cdot\xi_{2}%
^{\prime}}{z_{3}-z_{2}}\right]  \left[  \frac{k_{1}\cdot\xi_{3}^{\prime}%
}{z_{1}-z_{3}}+\frac{k_{2}\cdot\xi_{3}^{\prime}}{z_{2}-z_{3}}\right]  \\
+\frac{\xi_{2}\cdot\xi_{3}^{\prime}}{\left(  z_{3}-z_{2}\right)  ^{2}}\left[
\frac{k_{1}\cdot\xi_{2}^{\prime}}{z_{1}-z_{2}}+\frac{k_{3}\cdot\xi_{2}%
^{\prime}}{z_{3}-z_{2}}\right]  \left[  \frac{k_{1}\cdot\xi_{3}}{z_{1}-z_{3}%
}+\frac{k_{2}\cdot\xi_{3}}{z_{2}-z_{3}}\right]  \\
+\frac{\xi_{2}^{\prime}\cdot\xi_{3}}{\left(  z_{3}-z_{2}\right)  ^{2}}\left[
\frac{k_{1}\cdot\xi_{2}}{z_{1}-z_{2}}+\frac{k_{3}\cdot\xi_{2}}{z_{3}-z_{2}%
}\right]  \left[  \frac{k_{1}\cdot\xi_{3}^{\prime}}{z_{1}-z_{3}}+\frac
{k_{2}\cdot\xi_{3}^{\prime}}{z_{2}-z_{3}}\right]  \\
+\frac{\xi_{2}^{\prime}\cdot\xi_{3}^{\prime}}{\left(  z_{3}-z_{2}\right)
^{2}}\left[  \frac{k_{1}\cdot\xi_{2}}{z_{1}-z_{2}}+\frac{k_{3}\cdot\xi_{2}%
}{z_{3}-z_{2}}\right]  \left[  \frac{k_{1}\cdot\xi_{3}}{z_{1}-z_{3}}%
+\frac{k_{2}\cdot\xi_{3}}{z_{2}-z_{3}}\right]  \\
+\frac{\xi_{2}\cdot\xi_{3}}{\left(  z_{3}-z_{2}\right)  ^{2}}\frac{\xi
_{2}^{\prime}\cdot\xi_{3}^{\prime}}{\left(  z_{3}-z_{2}\right)  ^{2}}%
+\frac{\xi_{2}\cdot\xi_{3}^{\prime}}{\left(  z_{3}-z_{2}\right)  ^{2}}%
\frac{\xi_{2}^{\prime}\cdot\xi_{3}}{\left(  z_{3}-z_{2}\right)  ^{2}},
\end{array}
\right\}  \label{la}%
\end{align}
The result in Eq.(\ref{la}) includes contraction terms from $\partial_{2}%
X_{2}\cdot\cdot$ and $\partial_{3}X_{3}\cdot\cdot$. By setting $\xi_{2}%
=\xi_{2}^{\prime}=L_{2}$ and $\xi_{3}=\xi_{3}^{\prime}=L_{3}$ in
Eq.(\ref{la}), we obtain
\begin{align}
A^{L_{2}L_{2};L_{3}L_{3}} &  =\int d^{4}ze^{-\Lambda f}\left\{
\begin{array}
[c]{c}%
\left[  \frac{k_{1}^{L_{2}}}{z_{1}-z_{2}}+\frac{k_{3}^{L_{2}}}{z_{3}-z_{2}%
}\right]  ^{2}\left[  \frac{k_{1}^{L_{3}}}{z_{1}-z_{3}}+\frac{k_{2}^{L_{3}}%
}{z_{2}-z_{3}}\right]  ^{2}\\
+\frac{4L_{2}\cdot L_{3}}{\left(  z_{3}-z_{2}\right)  ^{2}}\left[  \frac
{k_{1}^{L_{2}}}{z_{1}-z_{2}}+\frac{k_{3}^{L_{2}}}{z_{3}-z_{2}}\right]  \left[
\frac{k_{1}^{L_{3}}}{z_{1}-z_{3}}+\frac{k_{2}^{L_{3}}}{z_{2}-z_{3}}\right]  \\
+\frac{2\left(  L_{2}\cdot L_{3}\right)  ^{2}}{\left(  z_{3}-z_{2}\right)
^{4}}%
\end{array}
\right\}  \label{ai}\\
&  \simeq\int dz_{2}e^{-\Lambda f}\left\{
\begin{array}
[c]{c}%
\left(  0+\frac{2E^{2}}{M_{2}}\left(  \frac{\left(  1-\tau\right)  ^{3}}{\tau
}\right)  \left(  z_{2}-\tilde{z}_{2}\right)  +\cdots\right)  ^{2}\left(
0-\frac{2E^{2}}{M_{3}}\frac{\left(  1-\tau\right)  ^{2}}{\tau}\left(
z_{2}-\tilde{z}_{2}\right)  +\cdots\right)  ^{2}\\
+\frac{4\left(  \frac{2E^{2}\tau}{M_{2}M_{3}}\right)  }{\left(  1-\frac
{1}{1-\tau}\right)  ^{2}}\left(  0+\frac{2E^{2}}{M_{2}}\left(  \frac{\left(
1-\tau\right)  ^{3}}{\tau}\right)  \left(  z_{2}-\tilde{z}_{2}\right)
+\cdots\right)  \\
\cdot\left(  0-\frac{2E^{2}}{M_{3}}\frac{\left(  1-\tau\right)  ^{2}}{\tau
}\left(  z_{2}-\tilde{z}_{2}\right)  +\cdots\right)  \\
+\frac{2\left(  \frac{2E^{2}\tau}{M_{2}M_{3}}\right)  ^{2}}{\left(  1-\frac
{1}{1-\tau}\right)  ^{4}}%
\end{array}
\right\}  .\label{ai2}%
\end{align}

In Eq.(\ref{ai2}) we have done Taylor expansion about the saddle point
$\tilde{z}_{2}$ for the first two terms and have put $0$ for each naive energy
order amplitude \cite{ChanLee,ChanLee2,CHLTY2,CHLTY1}. Most importantly, the
last two terms of Eq.(\ref{ai2}) involve crossing contraction terms and will
be seen soon that they are indeed of the same energy order as the first term
in Eq.(\ref{ai2}) which we have done Taylor expansion about the saddle point
$\tilde{z}_{2}$ due to the naive energy order problem.

Finally, we perform the integration in Eq.(\ref{ai2}) to obtain%

\begin{align}
A^{L_{2}L_{2};L_{3}L_{3}}  &  \simeq\int dz_{2}e^{-\Lambda f}\left\{
\begin{array}
[c]{c}%
\left(  \frac{2E^{2}}{M_{2}}\right)  ^{2}\left(  \frac{2E^{2}}{M_{3}}\right)
^{2}\frac{\left(  1-\tau\right)  ^{10}}{\tau^{4}}\left(  z_{2}-\tilde{z}%
_{2}\right)  ^{4}\\
-4\left(  \frac{2E^{2}}{M_{2}}\right)  ^{2}\frac{2E^{2}}{M_{3}^{2}}%
\frac{\left(  1-\tau\right)  ^{7}}{\tau^{3}}\left(  z_{2}-\tilde{z}%
_{2}\right)  ^{2}\\
+2\frac{2E^{2}}{M_{2}^{2}}\frac{2E^{2}}{M_{3}^{2}}\frac{\left(  1-\tau\right)
^{4}}{\tau^{2}}%
\end{array}
\right\} \\
&  =e^{-\Lambda f_{0}}\left\{
\begin{array}
[c]{c}%
\left(  \frac{2E^{2}}{M_{2}}\right)  ^{2}\left(  \frac{2E^{2}}{M_{3}}\right)
^{2}\frac{\left(  1-\tau\right)  ^{10}}{\tau^{4}}\left(  \frac{-2}{1}\right)
^{2}\left(  \frac{-1}{2}\right)  \left(  \frac{-3}{2}\right)  \left(  \frac
{1}{f_{0}^{\prime\prime}}\right)  ^{2}\sqrt{\frac{2\pi}{f_{0}^{\prime\prime}}%
}\\
-4\left(  \frac{2E^{2}}{M_{2}}\right)  ^{2}\frac{2E^{2}}{M_{3}^{2}}%
\frac{\left(  1-\tau\right)  ^{7}}{\tau^{3}}\left(  \frac{-2}{1}\right)
\left(  \frac{-1}{2}\right)  \frac{1}{f_{0}^{\prime\prime}}\sqrt{\frac{2\pi
}{f_{0}^{\prime\prime}}}\\
+2\frac{2E^{2}}{M_{2}^{2}}\frac{2E^{2}}{M_{3}^{2}}\frac{\left(  1-\tau\right)
^{4}}{\tau^{2}}\sqrt{\frac{2\pi}{f_{0}^{\prime\prime}}}%
\end{array}
\right\} \\
&  =e^{-\Lambda f_{0}}\frac{2E^{2}}{M_{2}^{2}}\frac{2E^{2}}{M_{3}^{2}}%
\frac{\left(  1-\tau\right)  ^{4}}{\tau^{2}}\sqrt{\frac{2\pi}{f_{0}%
^{\prime\prime}}}.
\end{align}
which gives the ratio consistent with Eq.(\ref{44}). This completes the
saddle-point calculation of $HSSA$ in Eq.(\ref{44}). We conclude that for the
$4$-point $2$ tachyon $2$ tensor $HSSA$, the degree of stringy scaling
dim$\mathcal{M}_{2}=1$ is the same as that of $3$ tachyons $1$ tensor $HSSA$
dim$\mathcal{M}_{1}=1$ calculated previously. However, as we will see in the
next section that, for the higher point $HSSA$, dim$\mathcal{M}_{2}$ will be
in general smaller than dim$\mathcal{M}_{1}$.

\subsection{General mass levels}

In this subsection, we extend our previous result to two general mass levels
$M_{i}^{2}=2(N_{i}-1)$, $i=2,3$. We first define the $K$-vectors%
\begin{equation}
K_{i}=\sum_{j\neq i}\frac{-k_{j}}{z_{j}-z_{i}}=(K_{i}^{P_{i}},K_{i}^{L_{i}%
},K_{i}^{T_{i}}), \label{KK}%
\end{equation}
which will be important in our following discussion. Note that in the hard
scattering limit $K_{i}^{P_{i}}=K_{i}^{L_{i}}$. By setting $z_{1}%
=0,z_{4}=\infty$ in%
\begin{equation}
k_{12}f=\sum_{i>j}k_{ij}\ln\left(  z_{i}-z_{j}\right)  ,
\end{equation}
we obtain%
\begin{equation}
k_{12}f=k_{12}\ln z_{2}+k_{13}\ln z_{3}+k_{23}\ln\left(  z_{3}-z_{2}\right)  .
\end{equation}

At the saddle point
\begin{align}
k_{12}\frac{\partial f}{\partial z_{2}}  &  =\frac{k_{12}}{z_{2}}-\frac
{k_{23}}{z_{3}-z_{2}}=M_{2}K_{2}^{L_{2}}=0,\\
k_{12}\frac{\partial f}{\partial z_{3}}  &  =\frac{k_{13}}{z_{3}}+\frac
{k_{23}}{z_{3}-z_{2}}=M_{3}K_{3}^{L_{3}}=0.
\end{align}
After setting $z_{3}=1$, we get%
\begin{align}
K_{2}^{L_{2}}  &  =-\frac{2E^{2}}{M_{2}}\frac{\left(  1-\tau\right)  ^{3}%
}{\tau}\left(  z_{2}-\tilde{z}_{2}\right)  ,\\
K_{3}^{L_{3}}  &  =\frac{2E^{2}}{M_{3}}\frac{\left(  1-\tau\right)  ^{2}}%
{\tau}\left(  z_{2}-\tilde{z}_{2}\right)  .
\end{align}

One important property of the $K_{i}$ vector is that its longitudinal
component $K_{i}^{L_{i}}$ vanishes at the saddle point%
\begin{equation}
\tilde{K}_{i}^{L_{i}}=0, \label{va}%
\end{equation}
The $K$-identities \cite{Regge, hard,Komaba} can then be written as%
\begin{equation}
\tilde{K}_{i}^{2}+2M_{i}\partial_{i}\tilde{K}_{i}^{L_{i}}=0. \label{K}%
\end{equation}
We will discuss these $K$-identities in section IV.

The $4$-point $HSSA$ with $2$ tachyons and $2$ tensors of Eq.(\ref{11}),
\begin{equation}
\left\vert p_{i},2m_{i},q_{i}\right\rangle =\left(  \alpha_{-1}^{T_{i}%
}\right)  ^{p_{i}}\left(  \alpha_{-1}^{L_{i}}\right)  ^{2m_{i}}\left(
\alpha_{-2}^{L_{i}}\right)  ^{q_{i}}\left\vert 0,k\right\rangle ,i=2,3
\end{equation}
with the mass levels%
\begin{equation}
N_{i}=p_{i}+2m_{i}+2q_{i}\text{ and }N=N_{2}+N_{3},
\end{equation}
can be calculated to be%
\begin{align}
A^{(p_{i},2m_{i},q_{i})}  &  =\int d^{4}z_{2}e^{-\Lambda f}\left(
K_{2}^{T_{2}}\right)  ^{p_{2}}\left(  K_{3}^{T_{3}}\right)  ^{p_{3}}\left(
-\partial_{2}K_{2}^{L_{2}}\right)  ^{q_{2}}\left(  -\partial_{3}K_{3}^{L_{3}%
}\right)  ^{q_{3}}\nonumber\\
&  \cdot\exp\left\{  \sum\limits_{j_{2}=1}^{2m_{2}}K_{2}\cdot\xi_{2,j_{2}%
}+\sum\limits_{j_{3}=1}^{2m_{3}}K_{3}\cdot\xi_{3,j_{3}}+\sum\limits_{j_{2}%
=1}^{2m_{2}}\sum\limits_{j_{3}=1}^{2m_{3}}\left[  \frac{\xi_{2,j_{2}}\cdot
\xi_{3,j_{3}}}{\left(  z_{3}-z_{2}\right)  ^{2}}\right]  \right\}
_{multilinear}\nonumber\\
&  =\int d^{4}ze^{-\Lambda f}\left(  K_{2}^{T_{2}}\right)  ^{p_{2}}\left(
K_{3}^{T_{3}}\right)  ^{p_{3}}\left(  -\partial_{2}K_{2}^{L_{2}}\right)
^{q_{2}}\left(  -\partial_{3}K_{3}^{L_{3}}\right)  ^{q_{3}}\nonumber\\
&  \sum\limits_{j=0}^{2m}j!C_{j}^{2m_{2}}C_{j}^{2m_{3}}\left[  \frac
{L_{2}\cdot L_{3}}{\left(  z_{3}-z_{2}\right)  ^{2}}\right]  ^{j}\left(
K_{2}^{L_{2}}\right)  ^{2m_{2}-j}\left(  K_{3}^{L_{3}}\right)  ^{2m_{3}-j}%
\end{align}
where in the last line we have set $\xi_{2,j}=L_{2},\xi_{3,j}=L_{3}$ and
$m=\min\left(  m_{2},m_{3}\right)  $. Plugging in the kenematics, we have%
\begin{align}
A^{(p_{i},2m_{i},q_{i})}  &  =\left(  \frac{4E^{2}\left(  1-\tau\right)  ^{3}%
}{\tau}\right)  ^{N_{2}/2+m_{2}+m_{3}}\left(  \frac{4E^{2}\left(
1-\tau\right)  }{\tau}\right)  ^{N_{3}/2}\sum\limits_{j=0}^{2m}j!C_{j}%
^{2m_{2}}C_{j}^{2m_{3}}\left[  \frac{-\tau}{2E^{2}\left(  1-\tau\right)  ^{3}%
}\right]  ^{j}\nonumber\\
&  \cdot\left(  \frac{-1}{2M_{2}}\right)  ^{2m_{2}+q_{2}}\left(  \frac
{-1}{2M_{3}}\right)  ^{2m_{3}+q_{3}}\int dz_{2}e^{-\Lambda f}\left(
z_{2}-\tilde{z}_{2}\right)  ^{2m_{2}+2m_{3}-2j}. \label{tau}%
\end{align}

We can now perform the integration to get%
\begin{align}
A^{(p_{i},2m_{i},q_{i})}  &  =e^{-\Lambda f_{0}}\left(  \frac{4E^{2}\left(
1-\tau\right)  ^{3}}{\tau}\right)  ^{N_{2}/2+m_{2}+m_{3}}\left(  \frac
{4E^{2}\left(  1-\tau\right)  }{\tau}\right)  ^{N_{3}/2}\sum\limits_{j=0}%
^{2m}j!C_{j}^{2m_{2}}C_{j}^{2m_{3}}\left[  \frac{-\tau}{2E^{2}\left(
1-\tau\right)  ^{3}}\right]  ^{j}\nonumber\\
&  \cdot\left(  \frac{-1}{2M_{2}}\right)  ^{2m_{2}+q_{2}}\left(  \frac
{-1}{2M_{3}}\right)  ^{2m_{3}+q_{3}}\left(  \frac{-2d}{df_{0}^{\prime\prime}%
}\right)  ^{m_{2}+m_{3}-j}\sqrt{\frac{2\pi}{f_{0}^{\prime\prime}}}.
\end{align}
After differentiation, we obtain%
\begin{align}
A^{(p_{i},2m_{i},q_{i})}  &  =e^{-\Lambda f_{0}}\left(  \frac{4E^{2}\left(
1-\tau\right)  ^{3}}{\tau}\right)  ^{N_{2}/2+m_{2}+m_{3}}\left(  \frac
{4E^{2}\left(  1-\tau\right)  }{\tau}\right)  ^{N_{3}/2}\sum\limits_{j=0}%
^{2m}j!C_{j}^{2m_{2}}C_{j}^{2m_{3}}\left[  \frac{-\tau}{2E^{2}\left(
1-\tau\right)  ^{3}}\right]  ^{j}\nonumber\\
&  \cdot\left(  \frac{-1}{2M_{2}}\right)  ^{2m_{2}+q_{2}}\left(  \frac
{-1}{2M_{3}}\right)  ^{2m_{3}+q_{3}}\sqrt{\frac{2\pi}{f_{0}^{\prime\prime}}%
}\frac{\left[  2\left(  m_{2}+m_{3}-j\right)  \right]  !}{2^{m_{2}+m_{3}%
-j}\left(  m_{2}+m_{3}-j\right)  !}\left(  f_{0}^{\prime\prime}\right)
^{-\left(  m_{2}+m_{3}-j\right)  }\nonumber\\
&  =\sqrt{\frac{2\pi}{f_{0}^{\prime\prime}}}\left(  \frac{4E^{2}\left(
1-\tau\right)  ^{3}}{\tau}\right)  ^{N_{2}/2}\left(  \frac{4E^{2}\left(
1-\tau\right)  }{\tau}\right)  ^{N_{3}/2}\left(  \frac{-1}{2M_{2}}\right)
^{2m_{2}+q_{2}}\left(  \frac{-1}{2M_{3}}\right)  ^{2m_{3}+q_{3}}\nonumber\\
&  \cdot\sum\limits_{j=0}^{2m}j!C_{j}^{2m_{2}}C_{j}^{2m_{3}}\left(  -2\right)
^{j}\frac{\left[  2\left(  m_{2}+m_{3}-j\right)  \right]  !}{\left(
m_{2}+m_{3}-j\right)  !}. \label{sum}%
\end{align}

Surprisingly, the summation in the last equation of Eq.(\ref{sum}) can be
calculated to be
\begin{equation}
\sum\limits_{j=0}^{2m}j!C_{j}^{2m_{2}}C_{j}^{2m_{3}}\left(  -2\right)
^{j}\frac{\left[  2\left(  m_{2}+m_{3}-j\right)  \right]  !}{\left(
m_{2}+m_{3}-j\right)  !}=\Gamma\left(  m_{2}+\frac{1}{2}\right)  \Gamma\left(
m_{3}+\frac{1}{2}\right)  ,
\end{equation}
and finally the $4$-point $2$ tensor $HSSA$ is%
\begin{align}
A^{(p_{i},2m_{i},q_{i})}  &  =e^{-\Lambda f_{0}}\sqrt{\frac{2\pi}%
{f_{0}^{\prime\prime}}}\left(  \frac{4E^{2}\left(  1-\tau\right)  ^{3}}{\tau
}\right)  ^{N_{2}/2}\left(  \frac{4E^{2}\left(  1-\tau\right)  }{\tau}\right)
^{N_{3}/2}\nonumber\\
&  \cdot\left(  \frac{-1}{2M_{2}}\right)  ^{2m_{2}+q_{2}}\left(  \frac
{-1}{2M_{3}}\right)  ^{2m_{3}+q_{3}}\Gamma\left(  m_{2}+\frac{1}{2}\right)
\Gamma\left(  m_{3}+\frac{1}{2}\right) \nonumber\\
&  =\sqrt{\frac{2\pi}{f_{0}^{\prime\prime}}}\left(  \frac{4E^{2}\left(
1-\tau\right)  ^{3}}{\tau}\right)  ^{N_{2}/2}\left(  \frac{4E^{2}\left(
1-\tau\right)  }{\tau}\right)  ^{N_{3}/2}\nonumber\\
&  \cdot\left[  \left(  \frac{1}{2}\right)  ^{m_{2}+q_{2}}\left(  \frac
{-1}{M_{2}}\right)  ^{2m_{2}+q_{2}}\left(  2m_{2}-1\right)  !!\right]  \left[
\left(  \frac{1}{2}\right)  ^{m_{3}+q_{3}}\left(  \frac{-1}{M_{3}}\right)
^{2m_{3}+q_{3}}\left(  2m_{3}-1\right)  !!\right]  .
\end{align}

One can now easily calculate the ratio%
\begin{align}
\frac{A^{\left(  \left\{  p_{2},p_{3}\right\}  ,2m_{2},2m_{3},q_{2}%
,q_{3}\right)  }}{A^{\left(  \left\{  N_{1},N_{2}\right\}  ,0,0\right)  }}  &
=\left[  \left(  \frac{1}{2}\right)  ^{m_{2}+q_{2}}\left(  -\frac{1}{M_{1}%
}\right)  ^{q_{2}+2m_{2}}\left(  2m_{2}-1\right)  !!\right] \nonumber\\
&  \cdot\left[  \left(  \frac{1}{2}\right)  ^{m_{3}+q_{3}}\left(  -\frac
{1}{M_{2}}\right)  ^{q_{3}+2m_{3}}\left(  2m_{3}-1\right)  !!\right]  ,
\end{align}
which is consistent with the calculation of $ZNS$ method in Eq.(\ref{22}).

\setcounter{equation}{0}

\section{The $n$-point two tensor $HSSA$}

In this section, we calculate the general $n$-point $HSSA$ with $n-2$ tachyons
and $2$ tensors of Eq.(\ref{11}). For the cases of $n\geq5$, one needs to
consider scattering processes out of the scattering plane with $r$-dimensional
transverse space. We begin with the case of $HSSA$ with $n-1$ tachyons and $1$
tensor. The general high energy states at each fixed mass level $M^{2}=2(N-1)$
can be written as \cite{Regge, hard}%
\begin{equation}
\left\vert \left\{  p_{i}\right\}  ,2m,q\right\rangle =\left(  \alpha
_{-1}^{T_{1}}\right)  ^{N+p_{1}}\left(  \alpha_{-1}^{T_{2}}\right)  ^{p_{2}%
}\cdots\left(  \alpha_{-1}^{T_{r}}\right)  ^{p_{r}}\left(  \alpha_{-1}%
^{L}\right)  ^{2m}\left(  \alpha_{-2}^{L}\right)  ^{q}\left\vert
0;k\right\rangle
\end{equation}
where $\sum_{i=1}^{r}p_{i}=-2(m+q)$ with\ $r\leq24$.

One generalizes the transverse polarization $e^{T}=(0,0,1)$ to $e^{\hat{T}%
}=(0,0,\vec{\omega})$ where%
\begin{equation}
\omega_{j}=\cos\theta_{j}\prod\limits_{\sigma=1}^{j-1}\sin\theta_{\sigma
}\text{, with }j=1,\cdots,r,\text{ }\theta_{r}=0 \label{ww}%
\end{equation}
are the solid angles in the transverse space spanned by $r$ transverse
directions $e^{T_{i}}$. For this case, the method of decoupling of $ZNS$ leads
to the ratios of $n$-point $HSSA$ \cite{Regge, hard,Komaba}%
\begin{equation}
\frac{A^{\left(  \left\{  p_{i}\right\}  ,2m,q\right)  }}{A^{\left(  \left\{
0_{i}\right\}  ,0,0\right)  }}=\frac{\left(  2m\right)  !}{m!}\left(
\frac{-1}{2M}\right)  ^{2m+q}\prod_{j=1}^{r}\omega_{j}^{p_{j}}\text{ (\# of
kinematic variables dependence reduced !).} \label{100}%
\end{equation}
These ratios are valid to all string loop orders, and for the string-tree
level case, Eq.(\ref{100}) can be rederived from the saddle-point calculation
\cite{Regge, hard,Komaba}.

We are now ready to discuss the $2$ tensor case. We first define%
\begin{equation}
L_{i}=\frac{k_{i}}{M_{i}},i=1,2.
\end{equation}
For simplicity of the amplitude calculation, we choose to put $2$ tensors at
$z_{1}$ and $z_{2}$. We consider $2$ high energy states of Eq.(\ref{11}) at
vertex $i$%
\begin{equation}
\left\vert 2m_{i}\right\rangle =\prod_{j=1}^{r}\left(  \alpha_{-1}^{T_{ij}%
}\right)  ^{p_{ij}}\left(  \alpha_{-1}^{L_{i}}\right)  ^{2m_{i}}\left(
\alpha_{-2}^{L_{i}}\right)  ^{q_{i}}\left\vert 0,k\right\rangle
\end{equation}
with the total mass level%
\begin{equation}
N_{i}=\sum_{j=1}^{r}p_{ij}+2m_{i}+2q_{i} \label{cc2}%
\end{equation}
where $i=1,2$ and $j=1\cdots r$.

The $K_{i}$ vectors can be similarly defined as in the $4$-point case in
Eq.(\ref{KK})%
\begin{align}
K_{1}  &  =\sum_{j\neq1}\frac{-k_{j}}{z_{j}-z_{1}}=(K_{1}^{P_{1}},K_{1}%
^{L_{1}},K_{1}^{T_{11}},K_{1}^{T_{12}},\cdots,K_{1}^{T_{1r}}),\\
K_{2}  &  =\sum_{j\neq2}\frac{-k_{j}}{z_{j}-z_{2}}=(K_{2}^{P_{2}},K_{2}%
^{L_{2}},K_{2}^{T_{21}},K_{2}^{T_{22}},\cdots,K_{2}^{T_{2r}}).
\end{align}
Since $K_{i}^{P_{i}}$ and $K_{i}^{L_{i}}$ vanishes at the saddle point as in
Eq.(\ref{va}), we can formally use two set of Eq.(\ref{ww}), $\omega_{1}^{j}$
and $\omega_{2}^{j}$ to parametrize the $K_{i}$ vectors
\begin{equation}
\tilde{K}_{i}^{T_{ij}}=\left\vert \tilde{K}_{i}\right\vert \omega_{i}%
^{j}\text{ , }i=1,2\text{ and }j\text{ }=1\cdots r. \label{cc1}%
\end{equation}
We stress that not all kinematic variables of $\omega_{1}^{j}$ and $\omega
_{2}^{j}$ are independent due to the various $K$-identities which put
constraints on $\omega_{i}^{j}$. We will discuss this issue in section IV.

For the kinematics of $n$-point amplitudes in the hard scattering limit,
$p=E\rightarrow\infty$, we define the $26$-dimensional momenta in the CM frame
to be%
\begin{align}
k_{1}  &  =\left(  E,-E,0^{r}\right)  ,\nonumber\\
k_{2}  &  =\left(  E,+E,0^{r}\right)  ,\nonumber\\
k_{j}  &  =\left(  -q_{j},-q_{j}\Omega_{1}^{j},\cdots-q_{j}\Omega_{25}%
^{j}\right)  \label{kkk}%
\end{align}
where $j=3,4,\cdots,n$, and%
\begin{equation}
\Omega_{i}^{j}=\cos\phi_{i}^{j}\prod\limits_{\sigma=1}^{i-1}\sin\phi_{\sigma
}^{j}\text{ with }\phi_{j-1}^{j}=0,\text{ }\phi_{i>r}^{j}=0\text{ and }%
r\leq\min\left\{  n-3,24\right\}  \label{k33}%
\end{equation}
are the solid angles in the $\left(  j-2\right)  $-dimensional spherical space
with $\sum_{i=1}^{j-2}\left(  \Omega_{i}^{j}\right)  ^{2}=1$. In
Eq.(\ref{kkk}), $0^{r}$ denotes the $r$-dimensional null vector. The condition
$\phi_{j-1}^{j}=0$ in Eq.(\ref{k33}) was chosen to fix the frame by using the
rotational symmetry. The independent kinematics variables can be chosen to be
$E$, some $\varphi_{i}^{j}$ and some fixed ratios of infinite $q_{j}$. See
Eq.(\ref{ex}) in section IV for the example of $n=5$.

The $HSSA$ of $n$-point string scattering with $2$ general tensors at $z_{2}$
and $z_{1}$\ can be expressed as%
\begin{equation}
A^{\left(  \left\{  p_{ij}\right\}  ,2m_{i},q_{i}\right)  }=\int_{0}%
^{1}dz_{n-2}\cdots\int_{0}^{z_{4}}dz_{3}\int_{0}^{z_{3}}dz_{2}\text{
}ue^{-\Lambda f}%
\end{equation}
where%
\begin{equation}
f=-\sum_{i>j}\frac{k_{i}\cdot k_{j}}{\Lambda}\ln\left(  z_{i}-z_{j}\right)
\text{, \ }\Lambda=-k_{1}\cdot k_{2}%
\end{equation}
and%
\begin{align}
u  &  =\left[  \prod_{j=1}^{r}\left(  K_{2}^{T_{2j}}\right)  ^{p_{2j}}\left(
K_{1}^{T_{1j}}\right)  ^{p_{1j}}\right]  \left(  -\partial_{2}K_{2}^{L_{2}%
}\right)  ^{q_{2}}\left(  -\partial_{1}K_{1}^{L_{1}}\right)  ^{q_{1}%
}\nonumber\\
&  \cdot\exp\left\{  \sum\limits_{j_{2}=1}^{2m_{2}}\left(  K_{2}\right)
\cdot\xi_{2,j_{2}}+\sum\limits_{j_{1}=1}^{2m_{1}}\left(  K_{1}\right)
\cdot\xi_{1,j_{1}}+\sum\limits_{j_{2}=1}^{2m_{2}}\sum\limits_{j_{1}=1}%
^{2m_{1}}\left[  \frac{\xi_{2,j_{2}}\cdot\xi_{1,j_{1}}}{\left(  z_{1}%
-z_{2}\right)  ^{2}}\right]  \right\}  _{multilinear}.
\end{align}
In the second line of the above equation, we can set $\xi_{2,k}=L_{2}%
,\xi_{1,k}=L_{1}$ and $m=\min\left(  m_{2},m_{1}\right)  $ to obtain%
\begin{align}
u  &  =\left(  -1\right)  ^{q_{1}+q_{2}}\sum\limits_{k=0}^{2m}\left[
\prod_{j=1}^{r}\left(  K_{2}^{T_{2j}}\right)  ^{p_{2j}}\left(  K_{1}^{T_{1j}%
}\right)  ^{p_{1j}}\right]  \left(  \partial_{2}K_{2}^{L_{2}}\right)  ^{q_{2}%
}\left(  \partial_{1}K_{1}^{L_{1}}\right)  ^{q_{1}}\nonumber\\
&  \cdot\left(  K_{2}^{L_{2}}\right)  ^{2m_{2}-k}\left(  K_{1}^{L_{1}}\right)
^{2m_{1}-k}k!C_{k}^{2m_{2}}C_{k}^{2m_{1}}\left[  \frac{k_{21}}{M_{2}%
M_{1}\left(  z_{1}-z_{2}\right)  ^{2}}\right]  ^{k}\nonumber\\
&  =\left(  -1\right)  ^{q_{1}+q_{2}}\sum\limits_{k=0}^{2m}u_{k}.
\end{align}

Since $\tilde{K}_{i}^{L_{i}}=0$, $i=1,2$ at the saddle-point, the leading
order term in each term of the summation is%
\begin{align}
&  \partial_{2}^{2m_{2}+2m_{1}-2k}\tilde{u}_{k}=C_{2m_{1}-k}^{2m_{2}%
+2mi_{1}-2k}\left[  \prod_{j=1}^{r}\left(  \tilde{K}_{2}^{T_{2j}}\right)
^{p_{2j}}\left(  \tilde{K}_{1}^{T_{1j}}\right)  ^{p_{1j}}\right]  \left(
\partial_{2}\tilde{K}_{2}^{L_{2}}\right)  ^{q_{2}+2m_{2}-k}\nonumber\\
&  \left(  \partial_{1}\tilde{K}_{1}^{L_{1}}\right)  ^{q_{1}}\left(
\partial_{2}\tilde{K}_{1}^{L_{1}}\right)  ^{2m_{1}-k}\left(  2m_{2}-k\right)
!\left(  2m_{1}-k\right)  !k!C_{k}^{2m_{2}}C_{k}^{2m_{1}}\left[  \frac{k_{21}%
}{M_{2}M_{1}\left(  \tilde{z}_{1}-\tilde{z}_{2}\right)  ^{2}}\right]  ^{k}%
\end{align}
where we have used%
\begin{equation}
\partial^{n}\left(  hg\right)  =\sum_{k=0}^{n}C_{k}^{n}h^{k}g^{n-k}.
\end{equation}

Now, in contrast to the $4$-point $HSSA$ calculation in Eq.(\ref{tau}) which
we know how to exactly solve the saddle-point $\tilde{z}_{2}=\frac{1}{1-\tau
}=\sec^{2}\frac{\phi}{2}$, for the general $n$-point $HSSA$ calculation, the
saddle-point $(\tilde{z}_{2},\tilde{z}_{3},\cdots,\tilde{z}_{n-2})$ may not be
solvable. So instead of solving the saddle-point equation, the next important
step is to introduce and apply the diagonal and the off-diagonal
$K$-identities%
\begin{equation}
\left\vert \tilde{K}_{i}\right\vert \left\vert \tilde{K}_{j}\right\vert
+2M_{i}\partial_{j}\tilde{K}_{i}^{L_{i}}=0\Rightarrow\left\{
\begin{array}
[c]{l}%
\tilde{K}_{i}^{2}+2M_{i}\partial_{i}\tilde{K}_{i}^{L_{i}}=0\text{, }i=j\\
\left\vert \tilde{K}_{i}\right\vert \left\vert \tilde{K}_{j}\right\vert
+\frac{2k_{ij}}{\left(  \tilde{z}_{i}-\tilde{z}_{j}\right)  ^{2}}=0\text{
}i\neq j
\end{array}
\right.  \label{KKK}%
\end{equation}
to obtain%
\begin{align}
\partial_{2}^{2m_{2}+2m_{1}-2k}\tilde{u}_{k}  &  =\left[  \prod_{j=1}%
^{r}\left(  \tilde{K}_{2}^{T_{2j}}\right)  ^{p_{2j}}\left(  \tilde{K}%
_{1}^{T_{1j}}\right)  ^{p_{1j}}\right]  \left(  -\frac{\tilde{K}_{2}^{2}%
}{2M_{2}}\right)  ^{q_{2}+2m_{2}-k}\nonumber\\
&  \left(  -\frac{\tilde{K}_{1}^{2}}{2M_{1}}\right)  ^{q_{1}}\left(
-\frac{\left\vert \tilde{K}_{2}\right\vert \left\vert \tilde{K}_{1}\right\vert
}{2M_{1}}\right)  ^{2m_{1}-k}\left(  2m_{2}+2m_{1}-2k\right)  !k!C_{k}%
^{2m_{2}}C_{k}^{2m_{1}}\left(  -\frac{\left\vert \tilde{K}_{2}\right\vert
\left\vert \tilde{K}_{1}\right\vert }{2M_{2}M_{1}}\right)  ^{k}. \label{key}%
\end{align}

One can now use Eq.(\ref{cc1}) and Eq.(\ref{cc2}) to get%
\begin{align}
\partial_{2}^{2m_{2}+2m_{1}-2k}\tilde{u}_{k}  &  =\left[  \prod_{j=1}%
^{r}\left(  \left\vert \tilde{K}_{2}\right\vert \omega_{2}^{j}\right)
^{p_{2j}}\left(  \left\vert \tilde{K}_{1}\right\vert \omega_{1}^{j}\right)
^{p_{1j}}\right]  \left(  -\frac{\tilde{K}_{2}^{2}}{2M_{2}}\right)
^{q_{2}+2m_{2}}\nonumber\\
&  \cdot\left(  -\frac{\tilde{K}_{1}^{2}}{2M_{1}}\right)  ^{q_{1}}\left(
-\frac{\left\vert \tilde{K}_{2}\right\vert \left\vert \tilde{K}_{1}\right\vert
}{2M_{1}}\right)  ^{2m_{1}}\left(  2m_{2}+2m_{1}-2k\right)  !k!C_{k}^{2m_{2}%
}C_{k}^{2m_{1}}\left(  -\frac{\tilde{K}_{2}^{2}}{2}\right)  ^{-k}\\
&  =\left[  \prod_{j=1}^{r}\left(  \omega_{2}^{j}\right)  ^{p_{2j}}\left(
\omega_{1}^{j}\right)  ^{p_{1j}}\right]  \tilde{K}_{2}^{N_{2}+2m_{2}+2m_{1}%
}\tilde{K}_{1}^{N_{1}}\left(  -\frac{1}{2M_{2}}\right)  ^{q_{2}+2m_{2}}\left(
-\frac{1}{2M_{1}}\right)  ^{q_{1}+2m_{1}}\nonumber\\
&  \cdot\left(  2m_{2}+2m_{1}-2k\right)  !k!C_{k}^{2m_{2}}C_{k}^{2m_{1}%
}\left(  -\frac{\tilde{K}_{2}^{2}}{2}\right)  ^{-k}. \label{diff}%
\end{align}

Finally, we can use the saddle-point method to evaluate the integral to obtain%
\begin{align}
A^{\left(  \left\{  p_{ij}\right\}  ,2m_{i},q_{i}\right)  }  &  \simeq\int
dz^{n-3}\text{ }\sum\limits_{k=0}^{2m}\frac{\left(  -1\right)  ^{q_{1}+q_{2}%
}\partial_{2}^{2m_{2}+2m_{1}-2k}\tilde{u}_{k}}{\left(  2m_{2}+2m_{1}%
-2k\right)  !}\left(  z_{2}-\tilde{z}_{2}\right)  ^{2m_{2}+2m_{1}%
-2k}e^{-\Lambda f}\nonumber\\
&  \simeq e^{-\Lambda f_{0}}\sum\limits_{k=0}^{2m}\frac{\left(  -1\right)
^{q_{1}+q_{2}}\partial_{2}^{2m_{2}+2m_{1}-2k}\tilde{u}_{k}}{\left(
2m_{2}+2m_{1}-2k\right)  !}\int_{0}^{1}dz_{2}\text{ }\left(  z_{2}-\tilde
{z}_{2}\right)  ^{2m_{2}+2m_{1}-2k}e^{-\frac{1}{4}\tilde{K}_{2}^{2}\left(
z_{2}-\tilde{z}_{2}\right)  ^{2}}\nonumber\\
&  =e^{-\Lambda f_{0}}\sum\limits_{k=0}^{2m}\frac{\left(  -1\right)
^{q_{1}+q_{2}}\partial_{2}^{2m_{2}+2m_{1}-2k}\tilde{u}_{k}}{\left(
2m_{2}+2m_{1}-2k\right)  !}\left(  -4\frac{d}{d\tilde{K}_{2}^{2}}\right)
^{m_{2}+m_{1}-k}\sqrt{\frac{4\pi}{\tilde{K}_{2}^{2}}}%
\end{align}
where $f_{0}=f(z_{2}=0,\tilde{z}_{3},\cdots,\tilde{z}_{n-2})$. After doing
differentiation and putting in Eq.(\ref{diff}), we obtain
\begin{align}
A^{\left(  \left\{  p_{ij}\right\}  ,2m_{i},q_{i}\right)  }  &  =e^{-\Lambda
f_{0}}\sum\limits_{k=0}^{2m}\frac{\left(  -1\right)  ^{q_{1}+q_{2}}%
\partial_{2}^{2m_{2}+2m_{1}-2k}\tilde{u}_{k}}{\left(  2m_{2}+2m_{1}-2k\right)
!}\sqrt{\frac{4\pi}{\tilde{K}_{2}^{2}}}\frac{\left(  2m_{2}+2m_{1}-2k\right)
!}{\left(  m_{2}+m_{1}-k\right)  !}\left(  \tilde{K}_{2}^{2}\right)
^{-\left(  m_{2}+m_{1}-k\right)  }\nonumber\\
&  =\left(  -1\right)  ^{q_{1}+q_{2}}e^{-\Lambda f_{0}}\sqrt{\frac{4\pi
}{\tilde{K}_{2}^{2}}}\tilde{K}_{2}^{N_{2}}\tilde{K}_{1}^{N_{1}}\left[
\prod_{j=1}^{r}\left(  \omega_{2}^{j}\right)  ^{p_{2j}}\left(  \omega_{1}%
^{j}\right)  ^{p_{1j}}\right] \nonumber\\
&  \cdot\left(  -\frac{1}{2M_{2}}\right)  ^{q_{2}+2m_{2}}\left(  -\frac
{1}{2M_{1}}\right)  ^{q_{1}+2m_{1}}\sum\limits_{k=0}^{2m}k!C_{k}^{2m_{2}}%
C_{k}^{2m_{1}}\left(  -2\right)  ^{k}\frac{\left[  2\left(  m_{2}%
+m_{1}-k\right)  \right]  !}{\left(  m_{2}+m_{1}-k\right)  !}.
\end{align}

We can now calculate the summation to get%
\begin{align}
A^{\left(  \left\{  p_{ij}\right\}  ,2m_{i},q_{i}\right)  }  &  =e^{-\Lambda
f_{0}}\sqrt{\frac{4}{\pi\tilde{K}_{2}^{2}}}\tilde{K}_{2}^{N_{2}}\tilde{K}%
_{1}^{N_{1}}\left[  \prod_{j=1}^{r}\left(  \omega_{2}^{j}\right)  ^{p_{2j}%
}\left(  \omega_{1}^{j}\right)  ^{p_{1j}}\right] \nonumber\\
&  \cdot\left(  \frac{1}{2}\right)  ^{q_{1}+q_{2}}\left(  -\frac{1}{M_{2}%
}\right)  ^{q_{2}+2m_{2}}\left(  -\frac{1}{M_{1}}\right)  ^{q_{1}+2m_{1}%
}\Gamma\left(  m_{2}+\frac{1}{2}\right)  \Gamma\left(  m_{1}+\frac{1}%
{2}\right) \nonumber\\
&  =e^{-\Lambda f_{0}}\sqrt{\frac{4\pi}{\tilde{K}_{2}^{2}}}\tilde{K}%
_{2}^{N_{2}}\tilde{K}_{1}^{N_{1}}\left[  \left(  \frac{1}{2}\right)
^{m_{1}+q_{1}}\left(  -\frac{1}{M_{1}}\right)  ^{q_{1}+2m_{1}}\left(
2m_{1}-1\right)  !!\prod_{j=1}^{r}\left(  \omega_{1}^{j}\right)  ^{p_{1j}%
}\right] \nonumber\\
&  \cdot\left[  \left(  \frac{1}{2}\right)  ^{m_{2}+q_{2}}\left(  -\frac
{1}{M_{2}}\right)  ^{q_{2}+2m_{2}}\left(  2m_{2}-1\right)  !!\prod_{j=1}%
^{r}\left(  \omega_{2}^{j}\right)  ^{p_{2j}}\right]  .
\end{align}
At last, the ratio can be easily calculated to be%
\begin{align}
\frac{A^{\left(  \left\{  p_{1j},p_{2j}\right\}  ,2m_{1},2m_{2},q_{1}%
,q_{2}\right)  }}{A^{\left(  \left\{  N_{1},N_{2}\right\}  ,0,0\right)  }}  &
=\left[  \left(  \frac{1}{2}\right)  ^{m_{1}+q_{1}}\left(  -\frac{1}{M_{1}%
}\right)  ^{q_{1}+2m_{1}}\left(  2m_{1}-1\right)  !!\prod_{j=1}^{r}\left(
\omega_{1}^{j}\right)  ^{p_{1j}}\right] \nonumber\\
&  \cdot\left[  \left(  \frac{1}{2}\right)  ^{m_{2}+q_{2}}\left(  -\frac
{1}{M_{2}}\right)  ^{q_{2}+2m_{2}}\left(  2m_{2}-1\right)  !!\prod_{j=1}%
^{r}\left(  \omega_{2}^{j}\right)  ^{p_{2j}}\right]  ,
\end{align}
which is consistent with the calculation of $ZNS$ method in Eq.(\ref{22}).

\setcounter{equation}{0}

\section{The $K$-identities and the degree of stringy scaling}

In the general $n$-point multi-tensor $HSSA$ calculation, one key step is to
introduce the diagonal and off-diagonal $K$-identities propsed in
Eq.(\ref{KKK}). The $K$-identities were originally proposed only for the
diagonal ones for the calculation of $n$-point $1$ tensor $HSSA$ calculation
\cite{Regge, hard,Komaba}. They are the direct results of the identification
of two different calculation of $HSSA$, namely, the saddle-point calculation
and the decoupling of zero-norm state ($ZNS$) calculation \cite{Regge,
hard,Komaba}.

We will show in this section that both diagonal and off-diagonal
$K$-identities for the $4$-point $HSSA$ with $1$ and $2$ tensor can be
analytically proved. For the $5$-point $HSSA$ calculation, although the
saddle-point equation is quadratic and can be exactly solved, the explicit
proof of the $K$-identities is too lengthy, and we will only prove them
numerically. We have also used maple to numerically prove the $K$-identities
for the $6$-point $HSSA$. Presumably, these $K$-identities are direct results
of conservation of momenta in the hard scattering limit \textit{at the
saddle-point} \cite{KID}.

It is important to point out that the diagonal $K$-identities put constraints
on the number of independent $\omega_{i}^{j}$, and the off-diagonal
$K$-identities put further constraints on the number of independent
$\omega_{i}^{j}$. So studying $K$-identities is crucial to calculate the
degree of stringy scaling dim$\mathcal{M}_{k}$.

\subsection{The $K$-identities}

We begin our discussion of the $K$-identities with one example of $6$-point
$1$ tensor $HSSA$ calculation. For $n=6$ and $r=3$, the ratio of $HSSA$ at
each mass level can be calculated by using saddle-point method to be
\cite{Regge, hard,Komaba}%
\begin{equation}
\frac{A^{\left(  p_{1},p_{2},p_{3},2m,q\right)  }}{A^{\left(  N,0,0,0\right)
}}=\frac{\left(  2m\right)  !}{m!}\left(  \frac{-1}{2M}\right)  ^{2m+q}%
\frac{\left(  \frac{\sum_{i\neq2,n=6}\frac{k_{i}^{T_{2}}}{\tilde{z}_{i}%
-\tilde{z}_{2}}}{\sum_{i\neq2,n=6}\frac{k_{i}^{T_{1}}}{\tilde{z}_{i}-\tilde
{z}_{2}}}\right)  ^{p_{2}}\left(  \frac{\sum_{i\neq2,n=6}\frac{k_{i}^{T_{3}}%
}{\tilde{z}_{i}-\tilde{z}_{2}}}{\sum_{i\neq2,n=6}\frac{k_{i}^{T_{1}}}%
{\tilde{z}_{i}-\tilde{z}_{2}}}\right)  ^{p_{3}}}{\left(  \frac{\sum_{i\neq
2,n}\frac{k_{i}^{T_{1}}}{\tilde{z}_{i}-\tilde{z}_{2}}}{\sqrt{2K\tilde{f}_{22}%
}}\right)  ^{2m+2q}}. \label{a22}%
\end{equation}
On the other hand, the decoupling of ZNS method gives \cite{Regge,
hard,Komaba}%
\begin{equation}
\frac{A^{\left(  p_{1},p_{2},2m,q\right)  }}{A^{\left(  N,0,0,0\right)  }%
}=\frac{\left(  2m\right)  !}{m!}\left(  \frac{-1}{2M}\right)  ^{2m+q}%
\omega_{1}^{p_{1}}\omega_{2}^{p_{2}}\omega_{3}^{p3}=\frac{\left(  2m\right)
!}{m!}\left(  \frac{-1}{2M}\right)  ^{2m+q}\frac{(\tan\theta_{1}\cos\theta
_{2})^{p_{2}}(\tan\theta_{1}\sin\theta_{2})^{p_{3}}}{(\cos\theta_{1})^{2m+2q}%
}. \label{b22}%
\end{equation}
Eq.(\ref{a22}) and Eq.(\ref{b22}) can be identified for any $p_{2}$, $p_{3}$,
$m$ and $q$ if%
\begin{align}
\sum_{i\neq2,n=6}\frac{k_{i}^{T_{1}}}{\tilde{z}_{i}-\tilde{z}_{2}}  &
=\sqrt{2\Lambda\tilde{f}_{22}}\cos\theta_{1}\text{,}\label{1a}\\
\text{ }\frac{\text{ }\sum_{i\neq2,n=6}\frac{k_{i}^{T_{2}}}{\tilde{z}%
_{i}-\tilde{z}_{2}}}{\text{ }\sum_{i\neq2,n=6}\frac{k_{i}^{T_{1}}}{\tilde
{z}_{i}-\tilde{z}_{2}}}  &  =\sqrt{2\Lambda\tilde{f}_{22}}\tan\theta_{1}%
\cos\theta_{2},\label{2b}\\
\frac{\text{ }\sum_{i\neq2,n=6}\frac{k_{i}^{T_{3}}}{\tilde{z}_{i}-\tilde
{z}_{2}}}{\text{ }\sum_{i\neq2,n=6}\frac{k_{i}^{T_{1}}}{\tilde{z}_{i}%
-\tilde{z}_{2}}}  &  =\sqrt{2\Lambda\tilde{f}_{22}}\tan\theta_{1}\sin
\theta_{2}. \label{3c}%
\end{align}
which allow us to formally express $\theta_{1}$ and $\theta_{2}$ defined in
Eq.(\ref{ww}) in terms of kinematics variables defined in Eq.(\ref{kkk}) as%
\begin{equation}
\theta_{1}=\arctan\frac{\sqrt{\left(  \tilde{k}_{\bot}^{T_{2}}\right)
^{2}+\left(  \tilde{k}_{\bot}^{T_{3}}\right)  ^{2}}}{\tilde{k}_{\perp}^{T_{1}%
}}\text{, }\theta_{2}=\arctan\frac{\tilde{k}_{\perp}^{T_{3}}}{\tilde{k}%
_{\perp}^{T_{2}}}. \label{RG}%
\end{equation}

It is important to note from Eq.(\ref{b22}) that the ratios of $6$-point
$HSSA$ with $r=3$ depends only on $2$ kinematics variables $\theta_{1}$ and
$\theta_{2}$ instead of $8$. We see that Eq.(\ref{1a}), Eq.(\ref{2b}) and
Eq.(\ref{3c}) gives the $K$-identity%
\begin{equation}
\tilde{K}_{2}^{2}+2M_{2}\partial_{2}\tilde{K}_{2}^{L_{2}}=0 \label{abc}%
\end{equation}
in Eq.(\ref{K}) for the case of $n=6$ and $r=3$. We note that the $K$-identity
puts $1$ constraint on $\omega_{1}$, $\omega_{2}$ and $\omega_{3}$. Thus there
are only $r-1=3-1=$ $2$ angles $\theta_{1}$ and $\theta_{2}$, and
dim$\mathcal{M}=8-2=6$.

For the second example of the $K$-identities, we consider $n$-point $2$ tensor
with general $r$ $HSSA$ calculated in section III. For this multi-tensor case,
in addition to the diagonal $K$-identities similar to Eq.(\ref{abc}), we
propose and include the off-diagonal $K$-identity in the second line of the
following equation%
\begin{equation}
\left\vert \tilde{K}_{i}\right\vert \left\vert \tilde{K}_{j}\right\vert
+2M_{i}\partial_{j}\tilde{K}_{i}^{L_{i}}=0\Rightarrow\left\{
\begin{array}
[c]{l}%
\tilde{K}_{i}^{2}+2M_{i}\partial_{i}\tilde{K}_{i}^{L_{i}}=0\text{, }i=1,2,\\
\left\vert \tilde{K}_{1}\right\vert \left\vert \tilde{K}_{2}\right\vert
+\frac{2k_{12}}{\left(  \tilde{z}_{1}-\tilde{z}_{2}\right)  ^{2}}=0,
\end{array}
\right.
\end{equation}
which we have used in Eq.(\ref{KKK}) in the saddle-point calculation of the
multi-tensor $HSSA$. The diagonal and off-diagonal $K$-identities proposed for
the calculation of $2$ tensor $HSSA$ are generalization of the diagonal
$K$-identity proposed for the calculation of $1$ tensor $HSSA$. They are the
key step to obtain the consistent multi-tensor $HSSA$ in the saddle-point calculation.

\subsection{The degree of stringy scaling}

For the general $n$-point $1$ tensor $HSSA$, it was shown that \cite{Regge,
hard,Komaba}
\begin{equation}
\text{dim}\mathcal{M}_{1}=\frac{\left(  r+1\right)  \left(  2n-r-6\right)
}{2}.
\end{equation}
For a given ($n,r$), we have \cite{Regge, hard,Komaba}%
\[%
\begin{tabular}
[c]{|c|c|c|c|c|}\hline
$\text{dim}\mathcal{M}_{1}$ & $r=1$ & $r=2$ & $r=3$ & $r=4$\\\hline
$n=4$ & $1$ &  &  & \\\hline
$n=5$ & $3$ & $3$ &  & \\\hline
$n=6$ & $5$ & $6$ & $6$ & \\\hline
$n=7$ & $7$ & $9$ & $10$ & $10$\\\hline
\end{tabular}
\ \ .\ \
\]

For the general $n$-point $2$ tensor $HSSA$, the off-diagonal $K$-identity in
Eq.(\ref{KKK}) puts $1$ further constraint except for the cases of $r=1$ which
contain only trivial $\omega_{1}^{1}=\omega_{2}^{1}=1$ according to
Eq.(\ref{ww}). We thus have%
\begin{align}
\dim M_{2}  &  =\frac{\left(  r+1\right)  \left(  2n-r-6\right)  }{2}-\left(
r-1\right)  +\left(  1-\delta_{r1}\right) \nonumber\\
&  =\dim M_{1}-\left(  r-1\right)  +\left(  1-\delta_{r1}\right)  \label{rr}%
\end{align}
where $\max r=\min\left(  n-3,24\right)  $. The $(r-1)$ factor in
Eq.(\ref{rr}) is the number of angles of $\omega_{2}^{j},$ $j=1,2,\cdots,r$,
and $\left(  1-\delta_{r1}\right)  $ factor is the effect of the off-diagonal
$K$-identity.

For a given ($n,r$), we have%
\[%
\begin{tabular}
[c]{|c|c|c|c|c|}\hline
$\text{dim}\mathcal{M}_{2}$ & $r=1$ & $r=2$ & $r=3$ & $r=4$\\\hline
$n=4$ & $1$ &  &  & \\\hline
$n=5$ & $3$ & $3$ &  & \\\hline
$n=6$ & $5$ & $6$ & $5$ & \\\hline
$n=7$ & $7$ & $9$ & $9$ & $8$\\\hline
\end{tabular}
\ \ \ \ .\ \
\]
We note that for $r=1,2$, $\dim\mathcal{M}_{2}=\dim\mathcal{M}_{1}$. For
$r\geq3$, $\dim\mathcal{M}_{2}=\dim\mathcal{M}_{1}-r+2$. It is interesting to
see that the off-diagonal $K$-identity contributes to the degree of scaling
dim$\mathcal{M}_{2}$ for $HSSA$ out of the scattering plane with $r\geq2$. We
see that for $HSSA$ with more than $2$ transverse directions or $r\geq3$, the
degree of scaling dim$\mathcal{M}_{2}$ decreases as one would have expected.

On the other hand, for a consistent check, for the case of $r=24$ and
$n\geq27$, one can show that $\dim\mathcal{M}_{2}\geq0$. For the case of
$r=n-3$ and $n\leq27$, we also have the consistent result
\begin{align}
\dim\mathcal{M}_{2}  &  =\frac{\left(  n-2\right)  \left(  n-3\right)  }%
{2}-\left(  n-3\right)  +\left(  2-\delta_{r1}\right) \nonumber\\
&  =\left(  \frac{n}{2}-2\right)  \left(  n-3\right)  +\left(  2-\delta
_{r1}\right)  \geq0.
\end{align}

For the general $n$-point $k$ tensor $HSSA$, it will be a challenge to
calculate dim$\mathcal{M}_{k}$. It is conjectured that for general higher $k$
tensor $HSSA$ calculation ($k\leq n$), the off-diaganal $K$-identities similar
to Eq.(\ref{KKK}) are crucial to show the consistent result that
dim$\mathcal{M}_{k}\geq0$.

\subsection{Checking the $K$-identities}

In this subsection, we will explicitly prove the $K$-identities of $HSSA$ for
the $4$-point $1$ and $2$ tensor cases. We will also discuss $K$-identities of
some higher point $HSSA$. We begin with the $4$-point kinematics in the hard
scattering limit

\bigskip%

\begin{align}
k_{1}  &  =\left(  E,-E,0\right)  ,\nonumber\\
k_{2}  &  =\left(  E,+E,0\right)  ,\nonumber\\
k_{3}  &  =\left(  -E,-E\cos\phi,-E\sin\phi\right)  ,\nonumber\\
k_{4}  &  =\left(  -E,E\cos\phi,E\sin\phi\right)  ,
\end{align}
which can be used to calculate%
\begin{equation}
k_{1}\cdot k_{2}=-2E^{2}\text{, \ \ }k_{2}\cdot k_{3}=2E^{2}\tau\text{,
\ \ }k_{1}\cdot k_{3}=2E^{2}\left(  1-\tau\right)  .
\end{equation}

For the $4$-point case, the saddle-point equation is linear, and the
saddle-point can be easily solved to be%
\begin{equation}
\tau=\sin^{2}\frac{\phi}{2}\text{, }\tilde{x}_{2}=\frac{1}{1-\tau}.
\end{equation}
For the $4$-point $1$ tensor case, we want to show that%
\begin{equation}
\tilde{K}_{2}^{2}+2M_{2}\partial_{2}\tilde{K}_{2}^{L_{2}}=0 \label{aaa}%
\end{equation}
where%
\begin{equation}
K_{2}\left(  z_{2}\right)  =\frac{k_{1}}{z_{2}}-\frac{k_{3}}{1-z_{2}}.
\end{equation}
In the hard scattering limit%
\begin{align}
\tilde{K}_{2}  &  =\left(  1-\tau\right)  \left(  k_{1}+\frac{k_{3}}{\tau
}\right)  ,\\
\tilde{K}_{2}^{2}  &  =\left(  1-\tau\right)  ^{2}\left(  k_{1}^{2}%
+\frac{k_{3}^{2}}{\tau^{2}}+2\frac{k_{1}\cdot k_{3}}{\tau}\right)  =\left(
1-\tau\right)  ^{2}\left(  0+\frac{0}{\tau^{2}}+\frac{4E^{2}\left(
1-\tau\right)  }{\tau}\right) \nonumber\\
&  =4E^{2}\frac{\left(  1-\tau\right)  ^{3}}{\tau}.
\end{align}
On the other hand%
\begin{equation}
\partial_{2}K_{2}\left(  z_{2}\right)  =-\frac{k_{1}}{z_{2}^{2}}-\frac{k_{3}%
}{\left(  1-z_{2}\right)  ^{2}},
\end{equation}
so%
\begin{equation}
\partial_{2}\tilde{K}_{2}=-\left(  1-\tau\right)  ^{2}\left(  k_{1}%
+\frac{k_{3}}{\tau^{2}}\right)
\end{equation}
and%
\begin{align}
\partial_{2}\tilde{K}_{2}^{L_{2}}  &  =-\left(  1-\tau\right)  ^{2}\left(
k_{1}+\frac{k_{3}}{\tau^{2}}\right)  \cdot\frac{k_{2}}{M_{2}}\nonumber\\
&  =-2E^{2}\frac{\left(  1-\tau\right)  ^{3}}{\tau}\frac{1}{M_{2}}.
\end{align}
The $K$-identity in Eq.(\ref{aaa}) is thus proved.

For the $4$-point $2$ tensor cases, we define%
\begin{equation}
K_{1}\left(  z_{2}\right)  =-\frac{k_{2}}{z_{2}-z_{1}}-\frac{k_{3}}%
{z_{3}-z_{1}}\text{, \ \ }K_{2}\left(  z_{2}\right)  =-\frac{k_{1}}%
{z_{1}-z_{2}}-\frac{k_{3}}{z_{3}-z_{2}}.
\end{equation}
It is straightforward to calculate%
\begin{equation}%
\begin{tabular}
[c]{|l|l|l|}\hline
& $\tilde{K}_{1}$ & $\tilde{K}_{2}$\\\hline
& $-k_{2}\left(  1-\tau\right)  -k_{3}$ & $\left(  k_{1}+\frac{k_{3}}{\tau
}\right)  \left(  1-\tau\right)  $\\\hline
$\partial_{1}$ & $-k_{2}\left(  1-\tau\right)  ^{2}-k_{3}$ & $k_{1}\left(
1-\tau\right)  ^{2}$\\\hline
$\partial_{2}$ & $k_{2}\left(  1-\tau\right)  ^{2}$ & $-\left(  k_{1}%
+\frac{k_{3}}{\tau^{2}}\right)  \left(  1-\tau\right)  ^{2}$\\\hline
\end{tabular}
\end{equation}
and%

\begin{equation}%
\begin{tabular}
[c]{|l|l|l|}\hline
& $\tilde{K}_{1}^{L_{1}}$ & $\tilde{K}_{2}^{L_{2}}$\\\hline
& $0$ & $0$\\\hline
$\partial_{1}$ & $\frac{-2E^{2}}{M_{1}}\left(  1-\tau\right)  \tau$ &
$-2E^{2}\left(  1-\tau\right)  ^{2}$\\\hline
$\partial_{2}$ & $-\frac{2E^{2}}{M_{1}}\left(  1-\tau\right)  ^{2}$ &
$-\left(  k_{1}+\frac{k_{3}}{\tau^{2}}\right)  \left(  1-\tau\right)  ^{2}%
$\\\hline
\end{tabular}
\ .
\end{equation}
One can now check the diagonal $K$-identities%

\begin{align}
&  \left\vert \tilde{K}_{1}\right\vert \left\vert \tilde{K}_{1}\right\vert
+2M_{1}\partial_{1}\tilde{K}_{1}^{L_{1}}=\left\vert \tilde{K}_{1}\right\vert
^{2}+2M_{1}\partial_{1}\tilde{K}_{1}^{L_{1}}\nonumber\\
&  =\left(  -k_{2}\left(  1-\tau\right)  -k_{3}\right)  ^{2}+2M_{1}\left(
-k_{2}\left(  1-\tau\right)  ^{2}-k_{3}\right)  \cdot\frac{k_{1}}{M_{1}%
}\nonumber\\
&  \symbol{126}2k_{2}\cdot k_{3}\left(  1-\tau\right)  -2\left(  k_{1}\cdot
k_{2}\left(  1-\tau\right)  ^{2}+k_{1}\cdot k_{3}\right) \nonumber\\
&  =4E^{2}\tau\left(  1-\tau\right)  -2\left(  -2E^{2}\left(  1-\tau\right)
^{2}+2E^{2}\left(  1-\tau\right)  \right)  =0
\end{align}
and%
\begin{align}
&  \left\vert \tilde{K}_{2}\right\vert \left\vert \tilde{K}_{2}\right\vert
+2M_{2}\partial_{2}\tilde{K}_{2}^{L_{2}}=\tilde{K}_{2}^{2}+2M_{2}\partial
_{2}\tilde{K}_{2}^{L_{2}}\nonumber\\
&  =\left(  k_{1}+\frac{k_{3}}{\tau}\right)  ^{2}\left(  1-\tau\right)
^{2}-2M_{2}\left(  k_{1}+\frac{k_{3}}{\tau^{2}}\right)  \cdot\frac{k_{2}%
}{M_{2}}\left(  1-\tau\right)  ^{2}\nonumber\\
&  \symbol{126}2k_{1}\cdot k_{3}\frac{\left(  1-\tau\right)  ^{2}}{\tau
}-2k_{1}\cdot k_{2}\left(  1-\tau\right)  ^{2}-2k_{2}\cdot k_{3}\frac{\left(
1-\tau\right)  ^{2}}{\tau^{2}}\nonumber\\
&  =\left(  1-\tau\right)  ^{2}\left[  \frac{4E^{2}\left(  1-\tau\right)
}{\tau}+4E^{2}-\frac{4E^{2}\tau}{\tau^{2}}\right]  =0.
\end{align}
Similarly, the off-diagonal $K$-identity is%
\begin{align}
&  \left\vert \tilde{K}_{1}\right\vert \left\vert \tilde{K}_{2}\right\vert
+2M_{1}\partial_{2}\tilde{K}_{1}^{L_{1}}=\left\vert \tilde{K}_{2}\right\vert
\left\vert \tilde{K}_{1}\right\vert +2M_{2}\partial_{1}\tilde{K}_{2}^{L_{2}%
}\nonumber\\
&  =\sqrt{\left(  -k_{2}\left(  1-\tau\right)  -k_{3}\right)  ^{2}}%
\sqrt{\left(  \left(  k_{1}+\frac{k_{3}}{\tau}\right)  \left(  1-\tau\right)
\right)  ^{2}}+2M_{1}\left(  1-\tau\right)  ^{2}\frac{k_{2}\cdot k_{1}}{M_{1}%
}\nonumber\\
&  \symbol{126}\sqrt{2k_{2}\cdot k_{3}\left(  1-\tau\right)  }\sqrt
{2k_{1}\cdot k_{3}\frac{\left(  1-\tau\right)  ^{2}}{\tau}}+2\left(
1-\tau\right)  ^{2}k_{1}\cdot k_{2}\nonumber\\
&  =\sqrt{4E^{2}\tau\left(  1-\tau\right)  }\sqrt{4E^{2}\left(  1-\tau\right)
\frac{\left(  1-\tau\right)  ^{2}}{\tau}}-4E^{2}\left(  1-\tau\right)  ^{2}=0.
\end{align}

We now turn to discuss the $5$-point $K$-identities. The $5$-point kinematics
can be written as%
\begin{align}
k_{1}  &  =\left(  \sqrt{p^{2}+M_{1}^{2}},-p,0,0\right)  ,\nonumber\\
k_{2}  &  =\left(  \sqrt{p^{2}+M_{2}^{2}},p,0,0\right)  ,\nonumber\\
k_{3}  &  =\left(  -\sqrt{q_{3}^{2}+M_{3}^{2}},-q_{3}\cos\phi_{32},-q_{3}%
\sin\phi_{32},0\right)  ,\nonumber\\
k_{4}  &  =\left(  -\sqrt{q_{4}^{2}+M_{4}^{2}},-q_{4}\cos\phi_{42},-q_{4}%
\sin\phi_{42}\cos\phi_{43},-q_{4}\sin\phi_{42}\sin\phi_{43}\right)
,\nonumber\\
k_{5}  &  =\left(  -\sqrt{q_{5}^{2}+M_{5}^{2}},-q_{5}\cos\phi_{52},-q_{5}%
\sin\phi_{52}\cos\phi_{53},-q_{5}\sin\phi_{52}\sin\phi_{53}\right)  .
\end{align}
The energy-momentum conservation laws imply%
\begin{align}
\sqrt{p^{2}+M_{1}^{2}}+\sqrt{p^{2}+M_{2}^{2}}  &  =\sqrt{q_{3}^{2}+M_{3}^{2}%
}+\sqrt{q_{4}^{2}+M_{4}^{2}}+\sqrt{q_{5}^{2}+M_{5}^{2}},\nonumber\\
0  &  =q_{3}\cos\phi_{32}+q_{4}\cos\phi_{42}+q_{5}\cos\phi_{52},\nonumber\\
0  &  =q_{3}\sin\phi_{32}+q_{4}\sin\phi_{42}\cos\phi_{43}+q_{5}\sin\phi
_{52}\cos\phi_{53},\nonumber\\
0  &  =0+q_{4}\sin\phi_{42}\sin\phi_{43}+q_{5}\sin\phi_{52}\sin\phi_{53}.
\end{align}
The energy conservation gives
\begin{equation}
2p+\left(  \frac{M_{1}^{2}}{2p}+\frac{M_{2}^{2}}{2p}\right)  +\cdots
=q_{3}+q_{4}+q_{5}+\left(  \frac{M_{3}^{2}}{2q_{3}}+\frac{M_{4}^{2}}{2q_{4}%
}+\frac{M_{5}^{2}}{2q_{5}}\right)  +\cdots,
\end{equation}
which in the hard scattering limit means%
\begin{equation}
2p=q_{3}+q_{4}+q_{5}.
\end{equation}

We assume%
\begin{equation}
q_{3}:q_{4}:q_{5}=1:r_{1}:r_{2}%
\end{equation}
to get%
\begin{equation}
q_{3}=\frac{2E}{1+r_{1}+r_{2}},q_{4}=\frac{2Er_{1}}{1+r_{1}+r_{2}},q_{5}%
=\frac{2Er_{2}}{1+r_{1}+r_{2}}.
\end{equation}
On the other hand, the momentum conservation laws become%
\begin{align}
0  &  =\cos\phi_{32}+r_{1}\cos\phi_{42}+r_{2}\cos\phi_{52},\nonumber\\
0  &  =\sin\phi_{32}+r_{1}\sin\phi_{42}\cos\phi_{43}+r_{2}\sin\phi_{52}%
\cos\phi_{53},\nonumber\\
0  &  =r_{1}\sin\phi_{42}\sin\phi_{43}+r_{2}\sin\phi_{52}\sin\phi_{53},
\end{align}
which can be used to solve $r_{2}$, $\phi_{53}$ and $\phi_{52}$. They are%
\begin{align}
&  r_{2}\left(  \phi_{32},\phi_{42},\phi_{43},r_{1}\right) \nonumber\\
&  =\sqrt{\left(  \cos\phi_{32}+r_{1}\cos\phi_{42}\right)  ^{2}+\left(
\sin\phi_{32}+r_{1}\sin\phi_{42}\cos\phi_{43}\right)  ^{2}+\left(  r_{1}%
\sin\phi_{42}\sin\phi_{43}\right)  ^{2}},\nonumber\\
&  =\sqrt{1+r_{1}^{2}+2r_{1}\left(  \cos\phi_{32}\cos\phi_{42}+\sin\phi
_{32}\sin\phi_{42}\cos\phi_{43}\right)  },
\end{align}%
\begin{align}
&  \phi_{53}\left(  \phi_{32},\phi_{42},\phi_{43},r_{1}\right) \nonumber\\
&  =\arctan\left(  \frac{r_{1}\sin\phi_{42}\sin\phi_{43}}{\sin\phi_{32}%
+r_{1}\sin\phi_{42}\cos\phi_{43}}\right)
\end{align}
and%
\begin{align}
&  \phi_{52}\left(  \phi_{32},\phi_{42},\phi_{43},r_{1}\right) \\
&  =\arctan\left(  \pm\sqrt{\frac{\left(  r_{1}\sin\phi_{42}\sin\phi
_{43}\right)  ^{2}+\left(  \sin\phi_{32}+r_{1}\sin\phi_{42}\cos\phi
_{43}\right)  ^{2}}{\left(  \cos\phi_{32}+r_{1}\cos\phi_{42}\right)  ^{2}}%
}\right)  .
\end{align}
Now, all kinematics variables can be expressed in terms of the following $5$
parameters%
\begin{equation}
E,\phi_{32},\phi_{42},\phi_{43},r_{1}. \label{ex}%
\end{equation}

The $5$-point kinematics in the hard scattering limit can thus be written as
(we take $r=2$ in Eq.(\ref{k33}))%
\begin{align}
k_{1}  &  =\left(  E,-E,0,0\right)  ,\nonumber\\
k_{2}  &  =\left(  E,E,0,0\right)  ,\nonumber\\
k_{3}  &  =\left(  -q_{3},-q_{3}\cos\phi_{32},-q_{3}\sin\phi_{32},0\right)
,\nonumber\\
k_{4}  &  =\left(  -q_{4},-q_{4}\cos\phi_{42},-q_{4}\sin\phi_{42}\cos\phi
_{43},-q_{4}\sin\phi_{42}\sin\phi_{43}\right)  ,\nonumber\\
k_{5}  &  =\left(  -q_{5},-q_{5}\cos\phi_{52},-q_{5}\sin\phi_{52}\cos\phi
_{53},-q_{5}\sin\phi_{52}\sin\phi_{53}\right)  ,
\end{align}
which can be used to calculate in the hard scattering limit%
\begin{align}
k_{12}  &  =-2E^{2},\nonumber\\
k_{13}  &  =\frac{2E^{2}}{1+r_{1}+r_{2}}\left(  1+\cos\phi_{32}\right)
,\nonumber\\
k_{23}  &  =\frac{2E^{2}}{1+r_{1}+r_{2}}\left(  1-\cos\phi_{32}\right)
,\nonumber\\
k_{24}  &  =\frac{2E^{2}r_{1}}{1+r_{1}+r_{2}}\left(  1-\cos\phi_{42}\right)
,\nonumber\\
k_{34}  &  =\frac{4E^{2}r_{1}r_{2}}{\left(  1+r_{1}+r_{2}\right)  ^{2}}\left(
-1+\cos\phi_{32}\cos\phi_{42}+\sin\phi_{32}\sin\phi_{42}\cos\phi_{43}\right)
.
\end{align}

We are now ready to discuss the $5$-point $K$-identities. We first discuss
$5$-point $1$ tensor $K$-identity. The saddle-point equation are%
\begin{align}
\frac{\partial f\left(  z_{2},z_{3}\right)  }{\partial z_{2}}  &  =0,\\
\frac{\partial f\left(  z_{2},z_{3}\right)  }{\partial z_{3}}  &  =0,
\end{align}
which can be expanded as%
\begin{align}
\frac{k_{12}}{z_{2}}+\frac{k_{23}}{z_{2}-z_{3}}+\frac{k_{24}}{z_{2}-1}  &
=0,\label{pp}\\
\frac{k_{13}}{z_{3}}+\frac{k_{23}}{z_{3}-z_{2}}+\frac{k_{34}}{z_{3}-1}  &  =0.
\label{qq}%
\end{align}
After some lengthy calculation, we end up with two quadratic equations for
$z_{2}$ and $z_{3}$%
\begin{align}
\left(  k_{12}+k_{24}+k_{23}\right)  \left(  k_{12}+k_{13}+k_{23}%
+k_{24}+k_{34}\right)  z_{2}^{2}  & \nonumber\\
-\left[  (k_{12}+k_{13}+k_{24}+k_{23}+k_{34})k_{12}+(k_{12}+k_{24}%
+k_{23})(k_{12}+k_{23}+k_{13})+k_{23}k_{34}\right]  z_{2}  & \nonumber\\
+(k_{12}+k_{23}+k_{13})k_{12}  &  =0,
\end{align}

\begin{align}
\left(  k_{13}+k_{34}+k_{23}\right)  \left(  k_{12}+k_{13}+k_{23}%
+k_{24}+k_{34}\right)  z_{3}^{2}  & \nonumber\\
-\left[  \left(  k_{12}+k_{13}+k_{24}+k_{23}+k_{34}\right)  k_{13}+\left(
k_{13}+k_{34}+k_{23}\right)  \left(  k_{13}+k_{23}+k_{12}\right)
+k_{23}k_{24}\right]  z_{3}  & \nonumber\\
+\left(  k_{12}+k_{23}+k_{13}\right)  k_{13}  &  =0.
\end{align}
Although the saddle-point equation is exactly solvable, it turns out to be too
lengthy to explicitly calculate it. We have instead successfully used maple to
check the $5$-point $1$ and $2$ tensor diagonal $K$-identities. The $5$-point
$2$ tensor off-diagonal $K$-identity can be similarly checked within the error
range. In addition, we have also numerically checked the validity of $6$-point
$1$ tensor diagonal $K$-identity.

\setcounter{equation}{0}

\section{Conclusion}

In this paper, we use the saddle-point method to calculate $n$-point $HSSA$
with $n-2$ tachyons and $2$ tensor states at arbitrary mass levels. During the
multi-tensor saddle-point calculation, one encounters the \textit{double naive
energy order problem}, and it turns out that one has to include contraction
terms from $\partial_{2}X_{2}\cdot\cdot$ and $\partial_{3}X_{3}\cdot\cdot$ in
order to achieve the consistent ratios calculated by the method of decoupling
of $ZNS$ in Eq.(\ref{44}).

More importantly, we discover the stringy scaling \cite{oleg} behavior of
these $HSSA$. For $HSSA$ with $r=1$,$2$, we found that dim$\mathcal{M}_{2}=$
dim$\mathcal{M}_{1}$. However, for $HSSA$ with more than $2$ transverse
directions or $r\geq3$, we found that the degree of stringy scaling
dim$\mathcal{M}_{2}$ of the $n-2$ tachyons and $2$ tensor $HSSA$ decreases
comparing to the degree of scaling dim$\mathcal{M}_{1}$ of the $n-1$ tachyons
and $1$ tensor $HSSA$ calculated previously.

In the saddle-point calculation of the general $n$-point multi-tensor $HSSA$,
the saddle-point $(\tilde{z}_{2},\tilde{z}_{3},\cdots,\tilde{z}_{n-2})$ may
not be exactly solvable. So instead of directly solving the saddle-point
equation for the $4$-point $HSSA$ calculation in Eq.(\ref{tau}), we propose a
set of diagonal and off-diagonal $K$-identities\ in Eq.(\ref{KKK}) to simplify
the $HSSA$ calculation in Eq.(\ref{key}) and thus demonstrate the stringy
scaling behavior of $HSSA$. Moreover, in section IV we explicitly prove both
the diagonal and off-diagonal $K$-identities for the $4$-point $HSSA$ and give
a numerical proof of these $K$-identities for the higher point $HSSA$.

For the general $n$-point $k$ tensor $HSSA$, we expect that there exist more
types of $K$-identities, and it will be a challenge to calculate the degree of
stringy scaling dim$\mathcal{M}_{k}$. It is conjectured that for general
higher $k$ tensor $HSSA$ calculation ($k\leq n$), the off-diaganal
$K$-identities similar to Eq.(\ref{KKK}) are crucial to show the consistent
result that dim$\mathcal{M}_{k}\geq0$.

Finally, there are evidences that these $K$-identities are direct results of
conservation of momenta in the hard scattering limit \textit{at the
saddle-point} \cite{KID}. Work in this direction is in progress.

\begin{acknowledgments}
We thank T. Yoneya and Y. Okawa for pointing out the importance of
$K$-identities in the early stage of this work. This work is supported in part
by National Science and Technology Council (NSTC) of Taiwan.
\end{acknowledgments}


\end{document}